\documentclass[preprint,12pt]{article}
\usepackage{epsfig}

\usepackage{amsmath,amssymb}
\usepackage{nameref,hyperref}
\usepackage{caption}

\def\d{\delta}

\def\a{\alpha}
\def\g{\gamma}

\def\s{\sigma}
\def\b{\beta}

\def\l{\lambda}

\usepackage{lineno}

\begin{document}

%\begin{frontmatter}



%\begin{flushleft}
{\Large
\textbf\newline{On the evolutionary emergence of predation} % Please use "sentence case" for title and headings (capitalize only the first word in a title (or heading), the first word in a subtitle (or subheading), and any proper nouns).
}
\newline
\\
Yaroslav Ispolatov\textsuperscript{1*},
Carlos Doebeli\textsuperscript{2},
Michael Doebeli\textsuperscript{3},
\\
\textbf{1} Departamento de F\'isica, Center for Interdisciplinary
Research in Astrophysics and Space Science, Universidad de Santiago de
Chile, Victor Jara  3493,  Santiago, Chile
\\
\textbf{2} Imperial College London, Department of Mathematics, South Kensington Campus, London SW7 2AZ, United Kingdom
\\
\textbf{3} Departments of Mathematics and Zoology, University of British
  Columbia, 6270 University Boulevard, Vancouver, B.C. V6T 1Z4, Canada
\\

% Use the asterisk to denote corresponding authorship and provide email address in note below.
* jaros007@gmail.com

\title{On the evolutionary emergence of predation}


% \author[1]{Yaroslav Ispolatov}
%   \affiliation[1]
%   {organization={Departamento de F\'isica, Center for Interdisciplinary
% Research in Astrophysics and Space Science, Universidad de Santiago de
% Chile},%Department and Organization
%             addressline={Victor Jara  3493}, 
%             city={Santiago},
%             country={Chile}}
% \author[2]{Carlos Doebeli}
% \affiliation[2] {organization={Imperial College London, Department of Mathematics, },%Department and Organization
%             addressline={South Kensington Campus}, 
%             city={London},
%             postcode={SW7 2AZ}, 
%             country={United Kingdom}}
%   \author[3]{Michael Doebeli}
% \affiliation[3] {organization={Departments of Mathematics and Zoology, University of British
%   Columbia},%Department and Organization
%             addressline={6270 University Boulevard}, 
%             city={Vancouver},
%             postcode={V6T 1Z4}, 
%             state={BC},
%             country={Canada}}  
          
\begin{abstract}
In models for the evolution of predation from initially purely
competitive species interactions, the propensity of predation is most often assumed to be a direct consequence of the relative morphological and physiological traits of interacting species. 
Here we explore a
model in which  predation ability is an independently evolving
phenotypic feature, so that even when the relative morphological or physiological traits allow for predation, predation only occurs if the predation ability of individuals has independently evolved to high enough values. 
In addition to delineating the conditions for the evolutionary emergence of predation, the model reproduces
stationary and non-stationary multilevel food webs 
with the top predators not necessarily having size superiority.
\end{abstract}



% \begin{keyword}
%  Evolution of predation%\sep Size superiority 
% %% keywords here, in the form: keyword \sep keyword

% %% PACS codes here, in the form: \PACS code \sep code
% \PACS 87.23.−n \sep 87.23.Kg \sep 87.18.−h
% %% MSC codes here, in the form: \MSC code \sep code
% %% or \MSC[2008] code \sep code (2000 is the default)

% \end{keyword}

%\end{frontmatter}

%\linenumbers

\section{Introduction}
\label{}
An important and possibly repeatedly occurring milestone in the evolution of
biological complexity is the emergence
of predation in various ecological niches and physical environments.
The very nature, chronology, and driving forces of such events are
still often shrouded in mystery \cite{sperling2013oxygen}.
There exists a somewhat mythological  concept of Garden of Eden or Shangri-La
\cite{bengtson2002origins} that describes the state of the biosphere
before the appearance of the first predators, where photosynthesizing and
chemotrophic cells peacefully floated, using available electromagnetic and chemical
energy, perhaps engaging in multilateral symbiosis and even competition,
but never killing each other for food consumption.  How long such a world could  have
existed (and whether it ever existed at all, \cite{de2009emergence})
still remains the subject of  debate. 

One reason for a high interest  in the evolution of predation is its
significance as one of the  strongest selective driver for evolution in other species \cite{bengtson2002origins}. 
Predation builds food webs, redistributes resources, both spatially
and metabolically, and potentially serves as an important driver of periods of
fast diversification and evolutionary expansion, such as the Cambrian
explosion \cite{erwin2011cambrian}.
Predators and their prey  may eventually form symbiotic relationships 
 and emerge as new organisms.
Explanations of several major transitions
in evolution (emergence of eukaryotes, sex, multicellularity, tissues,
active directed motility, etc) often cite emergence of predation
as a decisive factor \cite{bengtson2002origins}.
Predation is such an important part of life
that it has become one of the fundamental cultural metaphor, as it is
routinely used as a concept in economics % (``predatory pricing''), criminality (``sex predator'') 
or politics  %(``predatory democracy'') 
\cite{lyttkens1994predatory}.

The timing of the main evolutionary transitions in the history of
predation and the necessary and sufficient ecological conditions for
those transitions remain incompletely understood \cite{bengtson2002origins}.
In brief, the current view on those events can be
summarized as follows: It is universally acknowledged that early life,
perhaps based around hydrothermal vents, was too meagre and sparsely distributed
to sustain predation. Subsequent evolutionary invention of
photosynthesis led to widespread proliferation of stromatolites, or
bacterial mats, which could have been exposed to predation in the form of
parasitism from bacteriophages. The appearance of larger and more
complex eukaryotic cells is attributed to symbiotic arrangement
between former bacterial cells, which resulted from predatory acts of
prokaryotes engulfing or invading each other. The emergence of
multicellularity could have been a larger-size-favouring consequence of
an early arms race between predators and prey. The geologically
well-documented  and probably grazing-induced decline of bacterial
mats, the evolutionary emergence of protective hard mineralized exo- and
endo-skeletons and difficult-to-engulf macroscopic forms of life in
early pre-Cambrian times more than half a billion years ago,
 set the scene for the Cambrian burst of
evolutionary diversification and firmly established predation as the
main driver of further biological expansion.  

The emergence and subsequent diversification of predation appear to
require very special conditions, as these events
have to be preceded by the sufficient diversification and  biomass
accumulation of what will later become prey, and other prerequisites
created by the pre-predation biosphere, such as generation of sufficient oxygen
levels \cite{sperling2013oxygen}.  The evolutionary transition to
predation is also a risky one, as it is usually accompanied by phasing
out of the ability to utilize the normally steadily supplied primordial
energy resources,  such as light or fluxes of chemicals with usable
redox gradient. Furthermore, the paradigmatic
 requirement for a predator to overcome all types of prey defence
 for a successful attack (a kind of AND operation), compared to the
 sufficiency for prey protection of just a single functioning mode of defence
 (an OR operation), usually entails a high level of complexity for
 the predator.
However, once all the prerequisites are met
 and the first predators emerge, the ecological environment becomes
significantly more ''animated'', promoting rapid coevolution and
diversification of both predators and prey \cite{bengtson2002origins, bromham1998testing}.

Ecological and evolutionary systems with predator-prey 
interactions have not only long been a focus of paleontological and genetic studies, but also a subject of extensive theoretical
 modeling. The works of mathematicians and
physicists Alfred Lotka \cite{lotka1920analytical} and
Vito Volterra \cite{volterra1927fluctuations} in 1920s defined the namesake
system of equations, which has become a paradigm for the description of
ecological interactions between predators and prey, and the resulting
oscillatory population dynamics.  Subsequently, numerous models and
theoretical studies
(reviewed, for example, in \cite{abrams2000evolution}) have 
addressed the ecological and evolutionary properties of  food webs of
various complexity, stability, and function. 

An important achievement was reported  in 
\cite{loeuille2005evolutionary}, which studied the evolutionary auto-assembly of
a food web based on predator-prey and competitive interactions. It was shown that fairly 
elaborate and realistic-looking  food webs may evolve from a
single ancestor based on very simple ecological and evolutionary
rules: it is sufficient to postulate that selection acts on a single phenotypic
characteristic, body size, which controls both the intensity of predation
and the intensity of competition. 
The width of the resource distribution for prey and
the efficiency of conversion of the consumed food into predator offspring
were found to dramatically affect the emergent food-web structure and
functioning. Several later studies \cite{allhoff2016biodiversity,
  brannstrom2011emergence, 
  pillai2011metacommunity, bolchoun2017spatial, girardot2020does} further developed this
line of modeling, in particular allowing organisms to evolve multiple phenotypes affecting ecological interaction, while still assuming that both competition and predation are determined by the same phenotypes. 
A recent review \cite{fritsch2021identifying}
presents a comparative analysis of those models and finds a rather
strong dependence of the emergent food webs on the particular
assumptions made in each model, thus questioning the
 universality of the results obtained. It also stresses the
importance of conversion efficiency in food web evolution.

One particular feature in most existing models for the evolution of predation is that when the relative morphological or physiological phenotypes have the ''correct'' magnitude, predation is assumed to automatically occur. For example, if individuals have the ''right'' difference in size, then one (usually the larger) individual is automatically assumed to prey on the other (usually the smaller) individual. Thus,  a common assumption in
\cite{loeuille2005evolutionary,allhoff2016biodiversity,brannstrom2011emergence,
  pillai2011metacommunity,bolchoun2017spatial,girardot2020does,
  fritsch2021identifying} is
that a certain advantage in size (additive or multiplicative) of one
species over another is a necessary and sufficient condition for the
capability of a larger species to prey on a smaller one. 

However, there are many empirical examples
(e.g. \cite{sinclair2003patterns}) of ecological scenarios when
a larger body, while possibly saving one from becoming a prey,
does not necessarily lead to predation on
species of smaller individuals.  At various scales, a predatory species is much smaller than the prey (ranging from lytic phages attacking bacteria
\cite{hungate2021functional} to lions preying on elephants (e.g. \cite{john2009lion}).

%of 
%predators individually or collectively hunting species with
%body sizes equal or exceeding  those of the predators. 

%Many living and extinct large animals are
%grazers or, as baleen whales, prey on species with body size orders of magnitude smaller
%than normally assumed for predation body-size ratios. Conversely,
%there are many examples that range from lytic phages or oblige bacterial predators attacking other bacteria
%\cite{hungate2021functional} to lions preying on elephants (e.g. \cite{john2009lion}) of 
%predators individually or collectively hunting species with
%body sizes equal or exceeding  those of the predators. 

Thus, predation may not be an automatic consequence of relative phenotypes, but rather a life style whose evolution requires the evolution of specific
characteristics that determine whether a species is capable of
predation, and what fraction of an individual's energy budget comes from
it. Those characteristics  include the ability to search for prey, chase, kill, and digest
it, which in their turn depend on specific sensory, locomotive,
metabolic, and many other capabilities. It therefore seems useful to consider extensions of traditional
models by assuming that,  if it occurs, predation is still determined
by relative morphological and physiological phenotypes, but the actual propensity to be a predator in the first place is an independently evolving trait (comprising all those characteristic just mentioned). This is the approach we take in this paper in order to study   the evolutionary onset and
subsequent diversification of predation. 

We  consider the evolution of 
 predation-enabling traits in their simplest form. Modeling evolution of
realistic multidimensional phenotypes is potentially very complex, as well as computationally
expensive \cite{doebeli_ispolatov2017}. Hence, we reduce these
predation-defining phenotypes to a single ``degree of predation''  
 trait, a quantitative  phenotypic coordinate that is independent of
 other traits. Those generally numerous other traits, which may but do not have to
 include the body mass, determine aspects of competition for
 resources and other ecological interactions. Once the predation
 emerges, its efficiency is also dependent on how some of those other traits
 of predator and prey relate to each other.
 
In the following, we model the 
evolution of the degree of predation, which is described by a real number $0\leq p\leq 1$, under the assumption that both competitive and predatory interactions, if and when they occur, are determined by a set of morphological and physiological phenotypes (such as size) that are different from the degree of predation $p$. 
The starting point are simple Lotka-Volterra population
dynamics in which competitive interactions are determined by 
multi-dimensional phenotypes $\mathbf{x}=(x_1,x_2,...)$ as in  \cite{doebeli_ispolatov2017}, and
predation is absent (i.e., all individuals in the evolving community have trait value $p=0)$). If in the course of evolution 
the 1-dimensional degree of predation reaches values $p>0$ coding for
predation in a given population, then the multi-dimensional comprising the overall morphology and physiology of individuals phenotype
and determining competition is also also assumed to determine the efficiency of predation of
that population on other populations in the ecosystem.

Thus, evolution occurs in a multi-dimensional phenotype space, in which one of the
phenotypic coordinates, the degree of predation $p$, determines to what extend
an individual is a  predator, while the other phenotypic components $x$
determine competitive as well as predatory interactions. 

\section{Materials and methods}
\subsection{Model}
We consider populations in a well-mixed environment where densities of
individuals and resources have no 
spatial dependence. Individuals 
sustain themselves consuming external resource (e.g.  light
or chemical energy) and/or 
preying on other individuals. We quantify the feeding preferences of
an individual by its continuously varying degree of predation $p$
and the degree of primary  resource consumption $r$, with $0 \leq p \leq 1$ and $0 \leq r
 \leq 1$. 
Assuming that it is impossible to excel 
both in predation and resource competition due to various physical, physiological 
and biochemical constraints, a reasonable
assumption would be that individuals with higher $p$ have lower $r$, and vice versa. That is, more predatory individuals are less efficient primary resource consumers.  For simplicity, we assume that a perfect competitor for resources with $r=1$ cannot act as a predator at all, while a perfect predator with $p=1$, cannot
compete for resource, $r=0$. Intermediate values $0<p<1$ and $0<r<1$
reflect the impossibility to perfect both abilities and are constrained
by a generally nonlinear tradeoff, 
\begin{align}
\label{tr}
  p^{\l}+r^{\l}=1,
\end{align}
which links an individual's resource consumption to its predation rate,
$r=r(p)$. 
In addition to their degree of predation $p$, individuals are 
characterized by other morphological and physiological characteristics that impinge on ecological interactions (such
as sensory, locomotive, metabolic, and other biochemical rates, as well as body mass and shape, etc.). These traits are summarized in a generally multi-dimensional quantitative phenotype 
$\mathbf{x}\in\mathbf{R}^d$, where $d$ is the the number of phenotypic coordinates other than the predation ability $p$. 

The total set of evolving coordinates $(\mathbf{x},p)$ define all
ecological interactions. While $p$ and $r(p)$ determine the relative propensity of an individual to be a predator or a consumer of primary resources, the strength of the ecological interactions occurring between two individuals are determined by the phenotypic components $\mathbf{x}$ and $\mathbf{x'}$ of interacting individuals. 

For competitive interactions, consumption for primary resources and the resulting competition between
consumers is implemented as the classical logistic competition model
(see e.g. \cite{doebeli2011adaptive}). Accordingly, the per capita birth
rate of an individual $(\mathbf{x},p)$ due to resource consumption is assumed to be proportional to its resource
 consumption preference $r=r(p)$, i.e., the per capita birth rate of
 consumer $(\mathbf{x},p)$ is equal to $r(p)\beta$, where the parameter $\beta$ is the
 intrinsic birth rate of pure consumers.
 The per capita death rate %(ocr reduction  in birth rate) 
 resulting
 from competition between $(\mathbf{x},p)$ and another  phenotype
 $(\mathbf{x'},p')$ is given by  $r(p) r(p')\a(\mathbf{x,x'})/K(\mathbf{x})$, where $\a(\mathbf{x,x'})$ is the competition kernel and $K(\mathbf{x})$
is the environmental carrying capacity. The competition kernel
$\a(\mathbf{x,x'})$ can be thought of as a measure of the competitive
impact due to resource consumption of an individual with phenotype
$(\mathbf{x'},0)$ on an individual with phenotype
$(\mathbf{x},0)$. Here we we make the usual assumption that the
strength of competition has a maximum at $\mathbf{x}=\mathbf{x'}$ and
declines with increasing distance
$\vert\mathbf{x}-\mathbf{x'}\vert$. Thus, competition is strongest
between similar phenotypes and decreases with increasing
dissimilarity. The carrying capacity $K(\mathbf{x})$ is proportional
to the equilibrium population size of a pure consumer population $N$
that is monomorphic for phenotype $\mathbf{x}$, $N=(\beta-\d) K(\mathbf{x})/\a(x,x)$

As for predation, 
which is also determined by the phenotypes  $\mathbf{x}$ and  $\mathbf{x'}$ of interacting individuals, we assume that attack of an individual with phenotype $(\mathbf{x}$ on an individual with phenotype  $\mathbf{x'}$ occurs at a rate 
$p \g(\mathbf{x,x'})$. For example, if $\mathbf{x}$ describes body size, the attack kernel $\g(\mathbf{x,x'})$ essentially describes how the success 
of predation depends on the relative size of the 
 predator and its prey. In general $\mathbf{x}$ comprises all the morphological and physiological traits that are relevant for predation, 
 and the mathematically simplest assumption is that the attack kernel $\g(\mathbf{x,x'})$ has a
single maximum at a given
difference between the predator and prey traits,
$\mathbf{x}-\mathbf{x'}=\mathbf{m}$. For example, a predator may need to be bigger
than its prey for a successful
attack, but such size advantages cannot be too large due to tradeoffs with other essential traits, such as locomotion and perseverance. In contrast, in many case, e.g. for parasites and pathogens, the predator needs to be smaller than the prey, but not too small due to tradeoffs with other life history traits.
In general, depending on the nature and definition of $\mathbf{x}$, the
optimal difference $\mathbf{m}$ can be positive for some components, negative for others, or it can be (close to) 0.
 
 %For example, for $\mathbf{x}$ (or one
 %of its components) being the sensory ability, the
 %attack kernel describes how the success rate of predation depends on
 %the sensory advantage of predator over its prey. 
 
 While certain types of
prey could be easy to catch and consume, they may provide rather less
nutritional value, and vice versa. Thus, the contribution to the
reproduction rate of a predator with with phenotype $\mathbf{x}$ per caught prey individual
$\mathbf{x}'$ is characterized by a function $\chi(\mathbf{x,x'})$. Traditionally, this function is simply given by
a  coefficient $\chi < 1$, which reflects the efficiency
of conversion of consumed prey into offspring. Here we follow this tradition, but we note that more general
forms of $\chi(\mathbf{x,x'})$ may add additional complexity to the resulting evolutionary arms race 
dynamics \cite{fritsch2021identifying, th2019ecology}.

For a predator individual $(\mathbf{x},p)$, the attack rate $p \g(\mathbf{x,x'})$ translates into a birth rate (via the conversion efficiency $\chi$), whereas for the prey individual (with phenotype $\mathbf{x}'$) it directly translates into a death rate.  In addition to death due to the competition and predation,
we assume that there is a constant per capita death rate $\d$ that
reflects an external, constant mortality and is
independent of the  phenotype and the biotic environment. This term is needed to make the model meaningful for pure predators.
% We note that under these assumptions, the quantity $(\beta-\d) K(\mathbf{x})$ is the equilibrium population size of a pure consumer population that is monomorphic for $(\mathbf{x},0)$ in the absence of predators. (In particular, $K(\mathbf{x})$ can be thought of as a property of phenotype $\mathbf{x}$).
We will specify the functional form of $\a(\mathbf{x,x'})$, $\g(\mathbf{x,x'})$, and $K(\mathbf{x})$ below.

At any given time the system is assumed to be populated by one or
several strains $s=1,\ldots,$  with phenotypes  $(\mathbf{x}_s,p_s)$. The
number of individuals in each strain is assumed very large, so we ignore
separate birth and death events and describe the population dynamics
in terms of  population densities $N_s(t)$ continuously changing  in time,
% \begin{align}
% \label{e1}
%   \frac{ d N_s (t)}{d t}=N_s(t)
%  \left\{
%   r(p_s)\left[\beta -\sum_{s'} N_{s'}(t) r(p_{s'})\a(x_s,x_{s'})/K(x_s) \right] - \d \right.\\
%   \nonumber 
%   +\left.\sum_{s'} N_{s'}(t)
%   \left[ p_s \chi \g(x_s,x_{s'}) - p_{s'} \g(x_{s'},x_s) \right] \right\}.
% \end{align}
\begin{align}
\label{e1}
  \frac{ d N_s}{d t}=N_s
 \left[
  r_s\left(\beta -\sum_{s'} N_{s'}r_{s'}\a_{ss'}/K_s \right) - \d 
  %\nonumber 
  +\sum_{s'} N_{s'}
  \left( p_s \chi \g_{ss'} - p_{s'} \g_{s's} \right) \right].
\end{align}
Here $\a_{ss'}=\a(\mathbf{x}_s,\mathbf{x}_{s'})$,
$\g_{ss'}=\g(\mathbf{x}_s,\mathbf{x}_{s'})$, $r_s=r(p_s)$, and $K_s=K(\mathbf{x}_s)$.
The first term on the right hand side describes intrinsic reproduction, and the second and third terms are the death rates due to
competition with all strains $s'$ (including the strain $s$ itself), as well as to extrinsic mortality. 
The fourth term is the
reproduction of strain $s$ due to predation on all available strains,
attenuated by the conversion
efficiency  $\chi$. Finally, the last term describes the mortality that is  due to falling prey to potentially any strain (again including the strain $s$ itself, i.e., cannibalism is possible).

For simulating evolutionary dynamics in our system, as specified in the next section, we assume that mutant strains with very small population size
$\bar N$ are periodically added to the ecosystem. 
A mutant's ancestral strain is chosen with a probability proportional to
the current population size of such strain, and the mutant phenotype
$(\mathbf{x'}, p')$ is chosen in the phenotypic vicinity of the chosen ancestral strain $(\mathbf{x},p)$, \cite{fritsch2021identifying}.
To keep the number of strains in the system from ever increasing, strains with population density below a certain threshold are
considered extinct.

\subsection{Simulations}
As a compromise between computational and visualization simplicity and
the system's ability to display complex evolutionary dynamics \cite{doebeli2013symmetric}, we consider
two-dimensional phenotypes $\mathbf{x}$ (i.e., $d=2$), so that the entire $(\mathbf{x},p)$-phenotype space
is three-dimensional.

As explained above, the functions $\g(\mathbf{x,x'})$
and $\a(\mathbf{x,x'})$ are assumed to depend on the difference between $\mathbf{x}$ and
$\mathbf{x'}$ and have a single maximum at a particular vector value of $\mathbf{x} - \mathbf{x'}$ (with that vector being 0 for the competition kernel $\a(\mathbf{x,x'})$ ). 
Here we use simple Gaussian forms for these functions,
 \begin{align}
 \label{Gauss}
  \a(\mathbf{x,x'})=&\exp\left[ -\frac {\sum_{i=1}^{d}(x_i-x'_i)^2}{2\sigma_{\a}^2}
            \right] \bigg/\left(2\pi\sigma_{\a}^2\right)^{d/2}\\
   \nonumber 
  \g(\mathbf{x,x'})=&\exp\left[- \frac {\sum_{i=1}^{d}(x_i-x'_i-m)^2}{2\sigma_{\g}^2}
                \right]\bigg/\left(2\pi\sigma_{\g}^2\right)^{d/2}
  %\\
  % \nonumber
  %  \e(x,y)=&\chi_{\e}\\
  %  \nonumber
  %   \g(x,x')=&\frac {\chi_{\b}}{\chi_{\g}} 
   \nonumber
  \end{align}
Note that the difference vector at which the attack rate $\g$ has a maximum is assumed to be of the form $\mathbf{m}=(m,...,m)$ for simplicity (and without loss of generality). Note also that by rescaling strain densities and time units, we can assume
that the maximal predation rate is equal to $\left(2\pi\sigma_{\g}^2\right)^{-d/2}$ . Likewise, we can assume that $\a(\mathbf{x,x})=\left(2\pi\sigma_{\a}^2\right)^{-d/2}$ for all $\mathbf{x}$, since any other constant can be absorbed into the carrying capacity function $K(\mathbf{x})$. 
To rule out structural instabilities that can occur when both the competition kernel and the carrying capacity have Gaussian form \cite{gyllenberg2005impossibility,pigolotti2010gaussian}, the carrying capacity function is chosen with a quartic exponent: 

\begin{align}
 \label{resource}
  K(\mathbf{x})= K_0\exp \left[- \frac {\sum_{i=1}^{d}x_i ^4}{4\sigma_K^4}  \right]
\end{align}
We note that with such a quartic carrying capacity, the resulting pure
competition model \cite{doebeli2013symmetric, doebeli_ispolatov2017,rubin2021evolution}, which is obtained by setting $p=0$ for all strains and not allowing any mutations in the $p$-direction, exhibits evolutionary diversification for any competition width $\sigma_{\a}$ in the Gaussian competition kernel. The parameter $\sigma_K$ of  $K(\mathbf{x})$  sets the phenotypic scale of the
model, i.e., the range of viable phenotypes $\mathbf{x}$. Without loss of generality we set $\sigma_K=1$.

Each simulation run is initiated with a single consumer species with
all its phenotypic coordinates being set to zero. The population dynamics (\ref{e1}) are integrated numerically via 4-order Runge-Kutta method. Once every
$\Delta t_{mut}\sim 1$ time units a new  mutant strain is introduced with
a small population $\bar N \sim 10^{-7}$. The mutant's  ancestor  is
chosen randomly with the probability proportional to the ancestor's population. The
difference between the mutant and ancestral phenotypes is taken from
an isotropic Gaussian
distribution with zero mean and a small standard deviation $\mu =
10^{-2}$. The mutation frequency $1/\Delta t_{mut}$ is 
significantly less than the birth rate ($\b=4$, see below) to ensure
that the ecological dynamics is equilibrated (if it does so at all)
before a new mutant is introduced.  
Strains with population below a small
threshold (usually $\bar N/2$) are considered to be extinct and are
eliminated.

To keep simulation size manageable, every $\Delta
t_{merge}\sim 500$ time units strains with close phenotypes (within distance  $\sim \mu$)
are clustered, i.e. combined into one strain with phenotype equal to
the weighted mean phenotype of the strains chosen to make up one
cluster, and with population size equal to the sum of the population
sizes of the clustered strains. In terms of competition and predation
with phenotypically well-separated strains 
those closely clumped strains behave as a single strain with or
without clustering. At the same time, if not clustered, the ``relative'' population dynamics of
closely clumped strains that eventually results in one strain winning
over would have  been slow, creating extremely long and computationally
costly  transients \cite{rubin2023adaptive}. The simulations were run
for $t_{final}=1.4 \times 10^{6}$ time units, so that apparent steady states were reached in almost all scenarios.
 
Due to the complexity of the model, it appears unfeasible to explore the entire parameter space. Instead, we model
how changes in key parameters that are expected to affect the
emergence of predation affect the evolutionary dynamics of the whole system. 
In particular, we start with the  ``reference values''
specified in the second column in Table 1 and vary $\chi$,
$\l$, $K_0$, and $\mathbf {m}$.
\begin{table}[h!]
  \begin{center}
    \caption{Values of parameters used in the simulations.}
    \label{tab:table1}
    \begin{tabular}{|l|c|r|} 
            \hline  
      \textbf{Parameter} & \textbf{Reference value} & \textbf{Range of variation}\\
      \hline
$\s_{\a}$&0.5&constant\\
     \hline
$\s_{\g}$& 0.5&constant\\
     \hline
$\b$&4&constant\\
      \hline  
$\d$&1&constant\\
      \hline
      $\chi$& 0.5&0.063 to 0.75\\
      \hline  
$\l$&1&0.9 to 1.1\\
      \hline  
 $K_0$&4&1 to 16\\
      \hline
      $\mathbf {m}$&(0.5, 0.5)&(0, 0) to (1,1)\\
      \hline  
    \end{tabular}
  \end{center}
\end{table}

\section{Results}

\subsection{Conversion efficiency}
\begin{figure}
\centering
\includegraphics[width=0.32\textwidth]{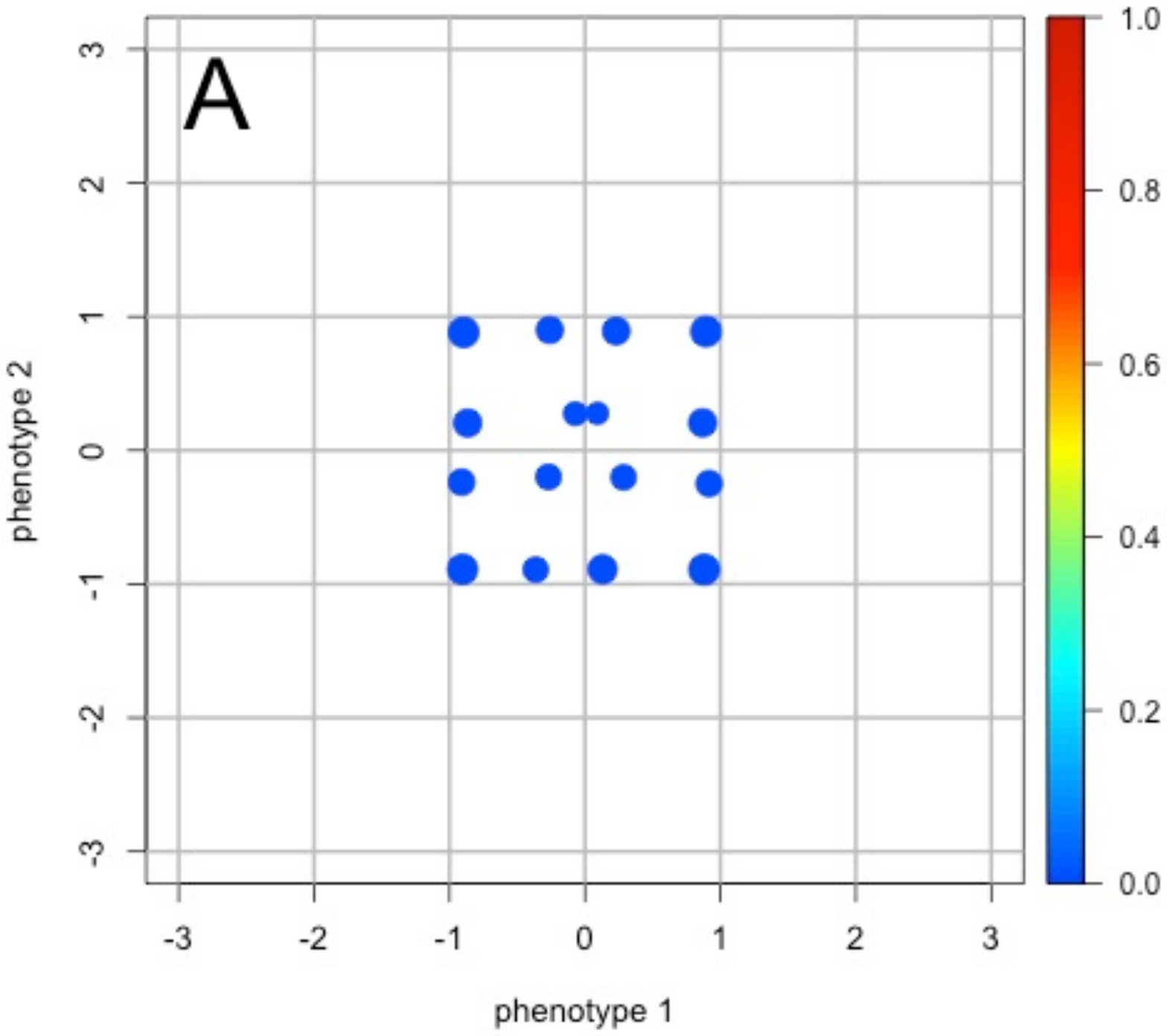}
\includegraphics[width=0.32\textwidth]{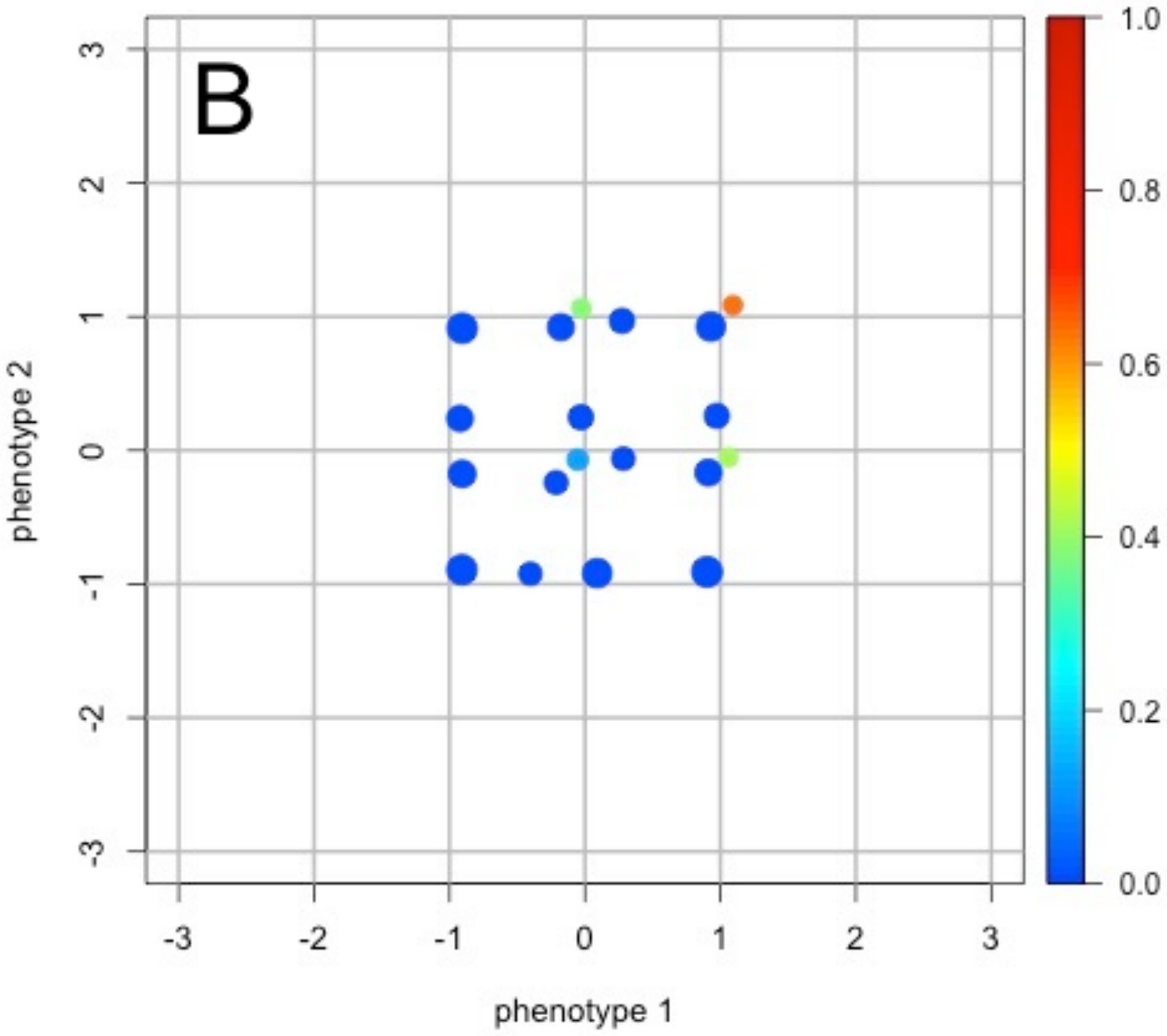}
\includegraphics[width=0.32\textwidth]{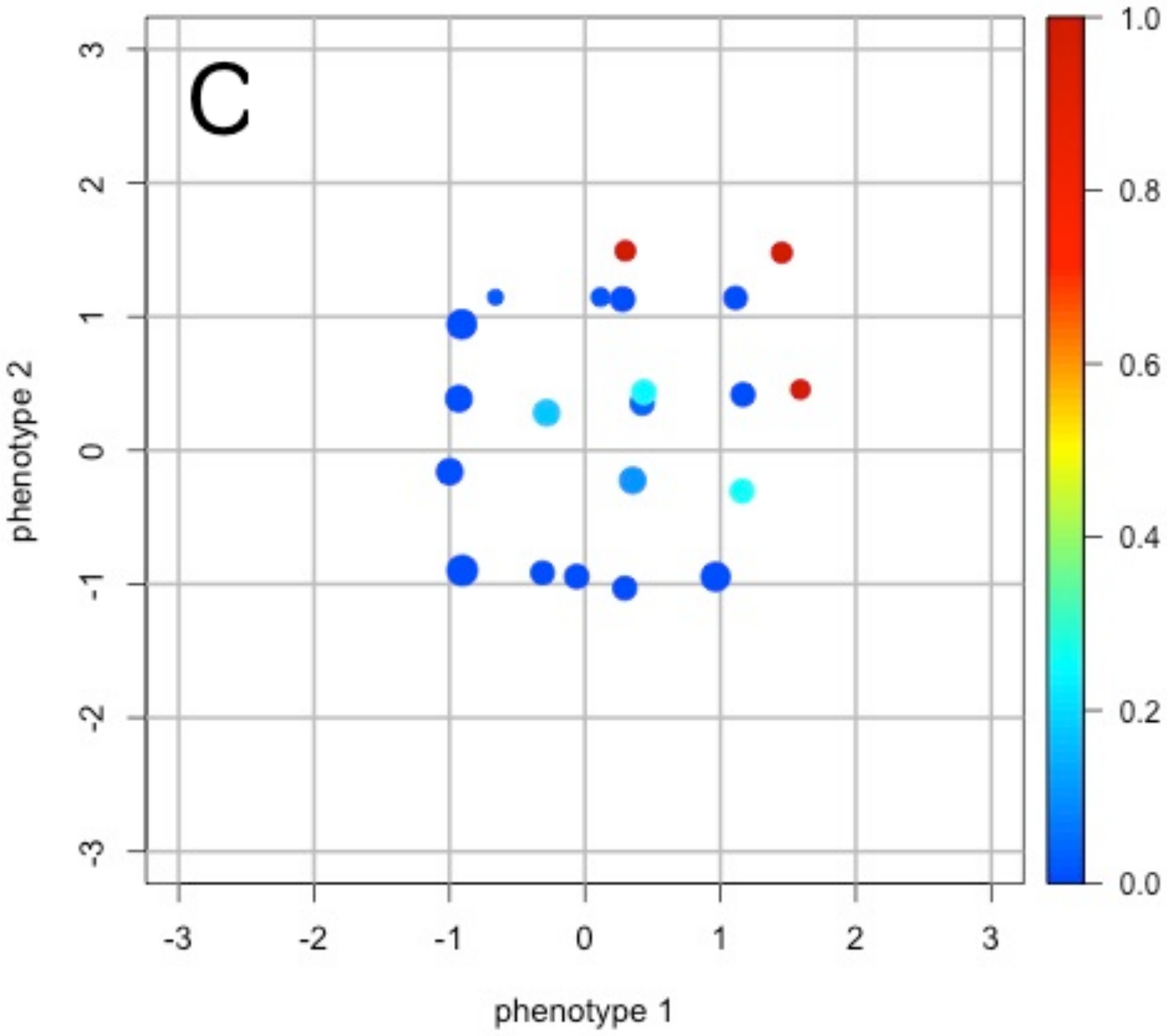}
\includegraphics[width=0.32\textwidth]{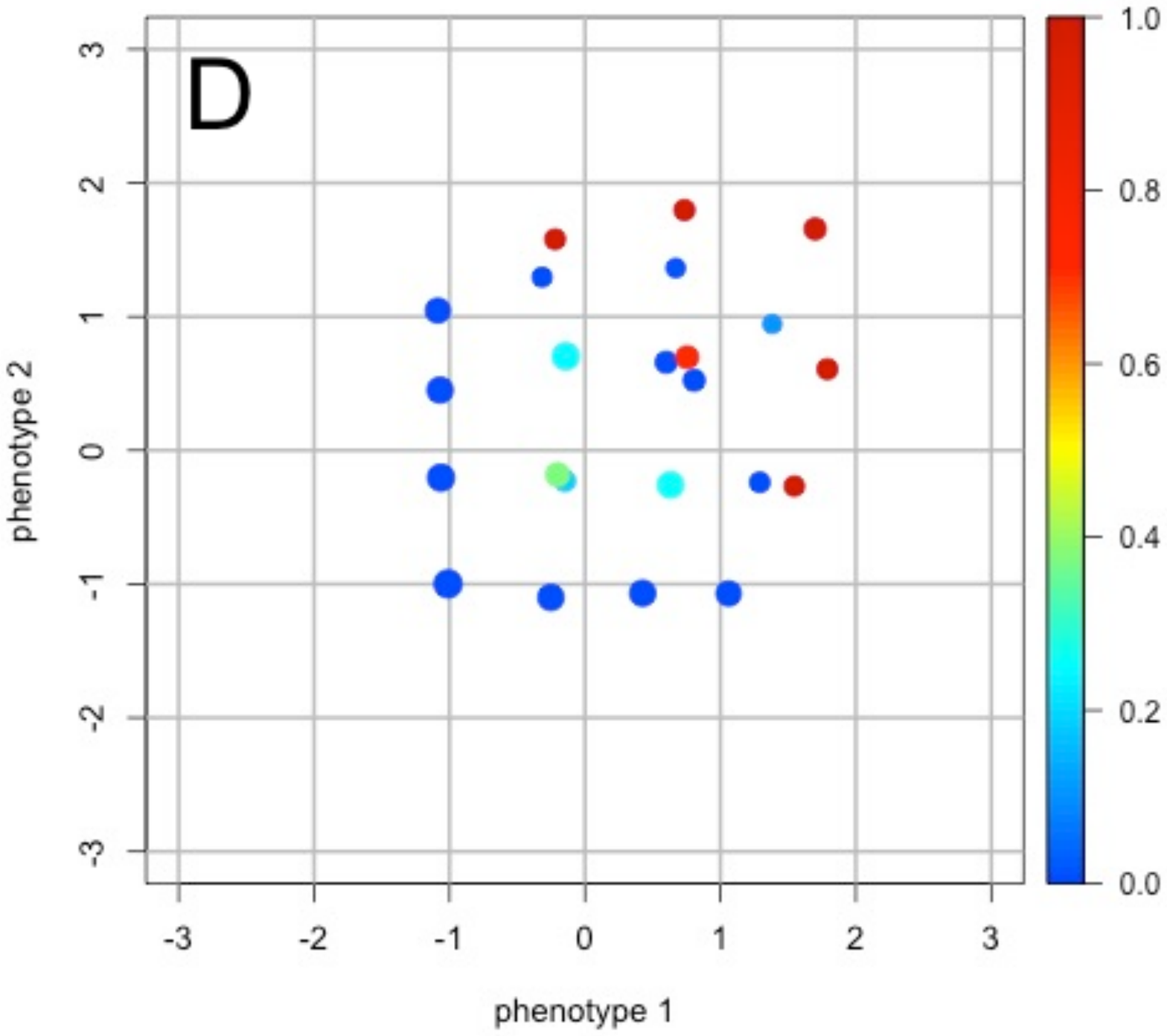}
\includegraphics[width=0.32\textwidth]{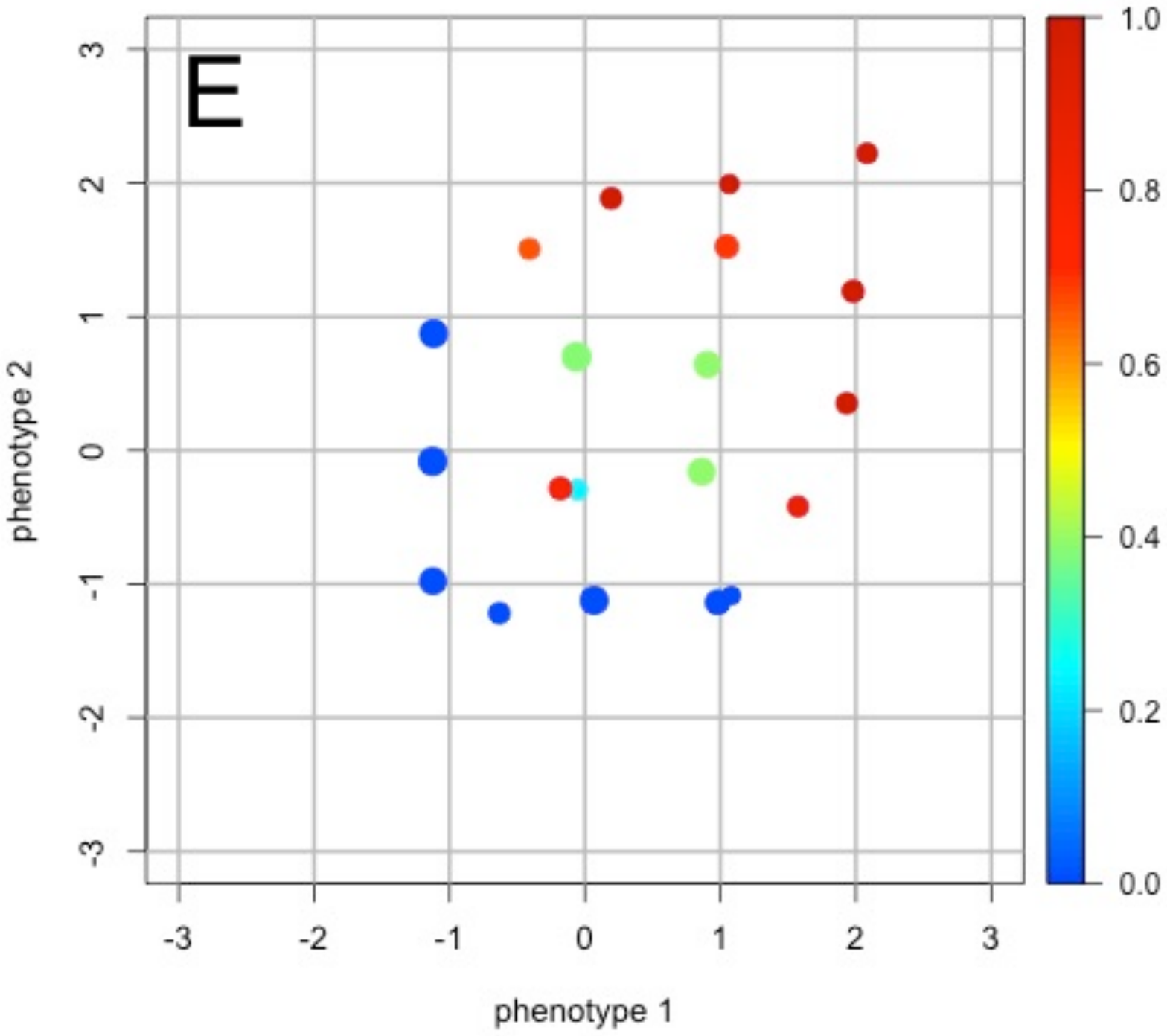}
\includegraphics[width=0.32\textwidth]{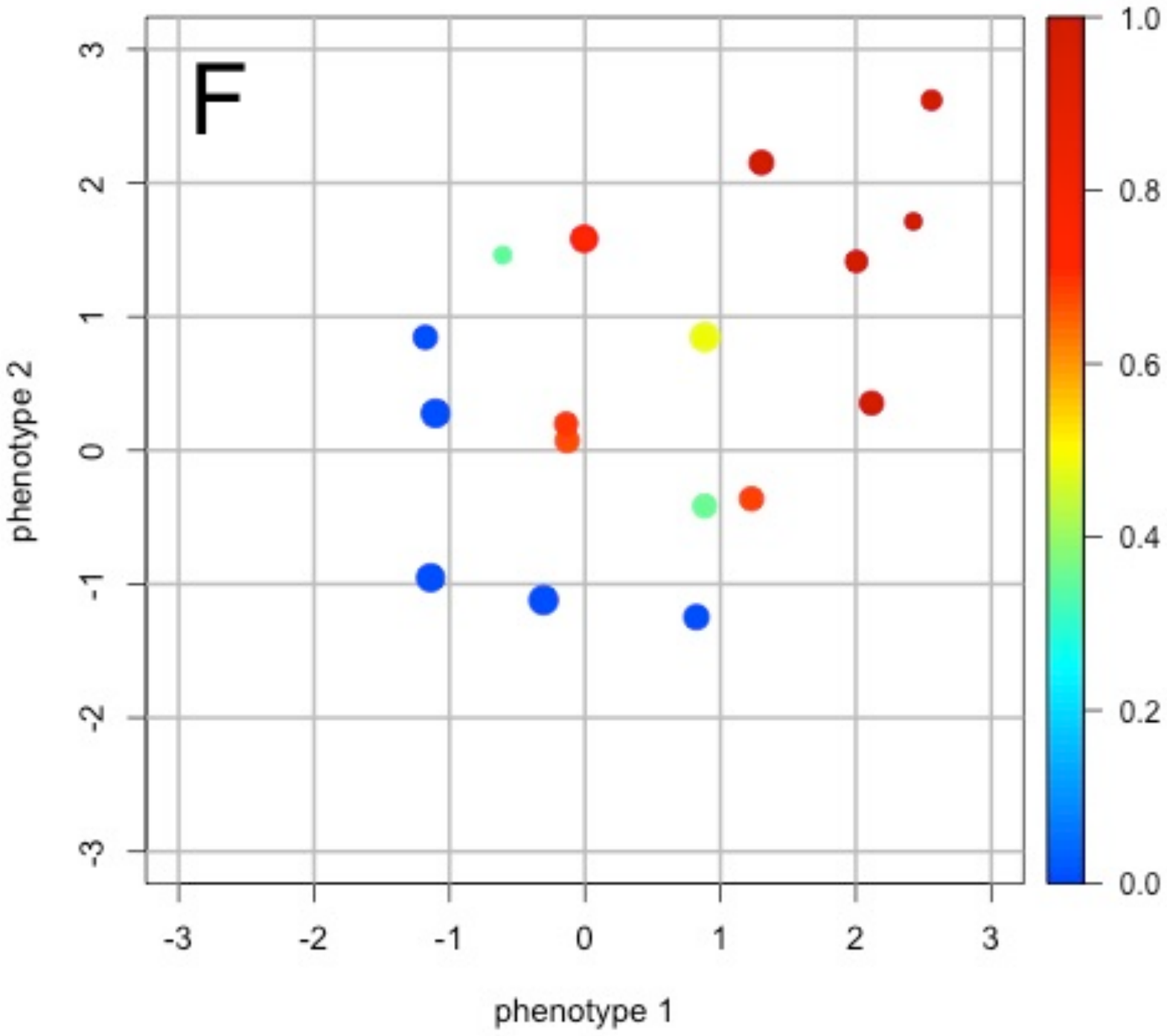}
 		\caption{ 
                  Snapshots of cluster distributions in 
                   communities after long-term evolution with
                   conversion efficiency (a) $\chi=0. 063$,
                   (b) $\chi=0. 125$,
                   (c) $\chi=0. 25$, 
                   (d) $\chi=0. 4$, 
                   (e) $\chi=0. 6$, 
                   (f) $\chi=0. 75$.
                   Here and in all subsequent snapshots and videos of
                   evolving systems, radii of 
                   circles are proportional to the square root of the
                   population density of the corresponding
                   species. The degree of predation
                   is indicated by the colour of the corresponding
                   circle, which varies from blue for $p=0$ to red for
                   $p=1$. The predator-prey relations between species
                   can be seen from the relative position of the corresponding circles:
                   here and in all subsequent snapshots and videos
                   except for Fig. 5a and Fig. 5b, a
                   predator's phenotypes $x_1$ and $x_2$ exceed 
                   those of its optimal prey by $m=0.5$.
                  Videos of the evolutionary processes that led to these 
                  configurations
                  can be found at
                  \href{https://figshare.com/s/1970d200680990011a96}{\underline{here}}. } 
		  \label{f1}
\end{figure}

It is intuitively clear and confirmed by existing models
\cite{fritsch2021identifying} that the conversion efficiency $\chi$ plays a
key role in emergence and formation of predator-prey foodwebs:
A very low prey conversion efficiency
restricts nutrient uptake from predation, and due to the 
tradeoff between predation and resource consumption (\ref{tr}), the
latter becomes the prime source of nutrients. Thus, it is expected
that for conversion efficiency below a certain threshold,  predatory
abilities do not evolve. Our model confirms this,
as shown in Fig. \ref{f1}a and the
corresponding video: for low $\chi$,  the well-studied  competition-driven evolutionary diversification \cite{doebeli_ispolatov2017} produces
a community of regularly spaced species. The regularity is due to the symmetries in the model, and the number of coexisting species is determined by the width of the competition kernel (16 species for $\sigma_{\a}=0.5$ in Fig. \ref{f1}a.)

A system  with twice larger conversion efficiency %, $\chi=0.125$,
initially also evolves by diversification of pure consumers. However, subsequently a single omnivore appears, which is
capable of both consuming resource and preying on several consumer species. Here and in the following,  species
with $p\leq1/3$ are classified as consumers, species with $1/3 < p \leq 2/3$ are
considered omnivores, while species with $2/3 < p $ are called
predators. In the example shown in Fig. \ref{f1}b and the corresponding video, two more species
develop omnivory before the evolving ecosystem reaches its steady state, but 
the conversion efficiency $\chi$ is still not high enough for 
pure predators to emerge.

Instead, pure predators only evolve when the predation efficiency is further increased, as shown in Fig. \ref{f1}c and the corresponding video. As for lower $\chi$, in such cases  the initial
stages of evolution consist 
of gradual diversification of consumers,  %(see the videos corresponding toFig. \ref{f1}a-c). 
and only when the diversity of consumers, which are potential prey, is sufficiently developed, 
it is possible for omnivores and then pure predators to evolve.  Note
that the evolution of predators does not necessarily result
in a reduction in the number of consumer species. On the contrary, the evolution of
predation can result in an increase in the diversity of prey (consumer) species (Fig. \ref{f1}c).

Other evolutionary scenarios can be seen for even higher predation efficiencies $\chi$. For example, in the case shown in Fig. \ref{f1}d and the corresponding video, the single initial consumer species does not diversify
into other consumers. Instead, it evolves increased phenotypes in both the  $x_1$ and $x_2$ direction while at the same
time evolving higher $p$, i.e., an increase in the share of predation in its energy budget. This evolutionary pattern is driven by
cannibalism, which requires a certain degree of predation. Thus, for a
sufficiently high predation efficiency, preying on the members of the same species becomes
feasible, and due to the preference of predators to attack a
prey with smaller $x_1,x_2$ (specified as the optimal offset $m$), the cannibalistic species
evolves towards larger $x$. In the Appendix we present
an adaptive dynamics explanation for this evolutionary pattern and
estimate the threshold $\chi$ for it to be possible. Subsequently, the cannibalism-driven
single species evolution becomes unstable with respect to
diversification, which produces sets of species that specialize in
resource consumption, predation, or omnivory (see Fig. \ref{f1}d and the corresponding video).

Evolutionary scenarios with even higher predation efficiencies are similar to the
one just described, but with larger  $\chi$ resulting in fewer 
 consumers and more predators (Figs. 1e, 1f). It is visible in Figs. 1e, 1f
that for large $\chi$, pure predators are either specializing in preying on consumers (the
predator species close to the center of phenotype space in Fig. 1e) or in preying on
other predators (predators with
larger $x_1,x_2$ visible in the upper right corner of phenotype space in Fig. 1e).
Thus, very efficient predation allows the system to evolve a
multilevel food chain, consisting of a community of consumers, primary,
and higher-order predators. We also observe signs of the well-known
dynamical complexity of multilevel predation food webs \cite{hastings1991chaos}: 
the ecological as well as the evolutionary dynamics in the system with the highest 
efficiencies that we studied were never observed to come
to a steady state.

Overall, the diversity in the entire evolving ecosystems (comprising consumers, omnivores, and pure predators) peaks
at some intermediate conversion efficiency ($\chi\sim 0.4$ for the
parameters used for Fig. 1), as shown in Fig. 2a. At the same time, the diversity of pure
predators  increases with  $\chi$ and saturates  at
$\chi\approx 0.6$.

The dynamics  of diversity and  population density  for a typical
evolutionary scenario with $\chi=0.5$ are shown in Figs 2b and
2c. Note that the total population density, as
well as sub-total population densities of all consumers, all omnivores, and all
predators, respectively,  come to steady state before the complete diversification of the
corresponding group of species. Thus, the sub-total equilibrium population densities of these three
``classes'' of species do not strongly depend on the precise number
of species in each class after a certain threshold diversification has been
reached. 

\begin{figure}
\centering
\includegraphics[width=0.32\textwidth]{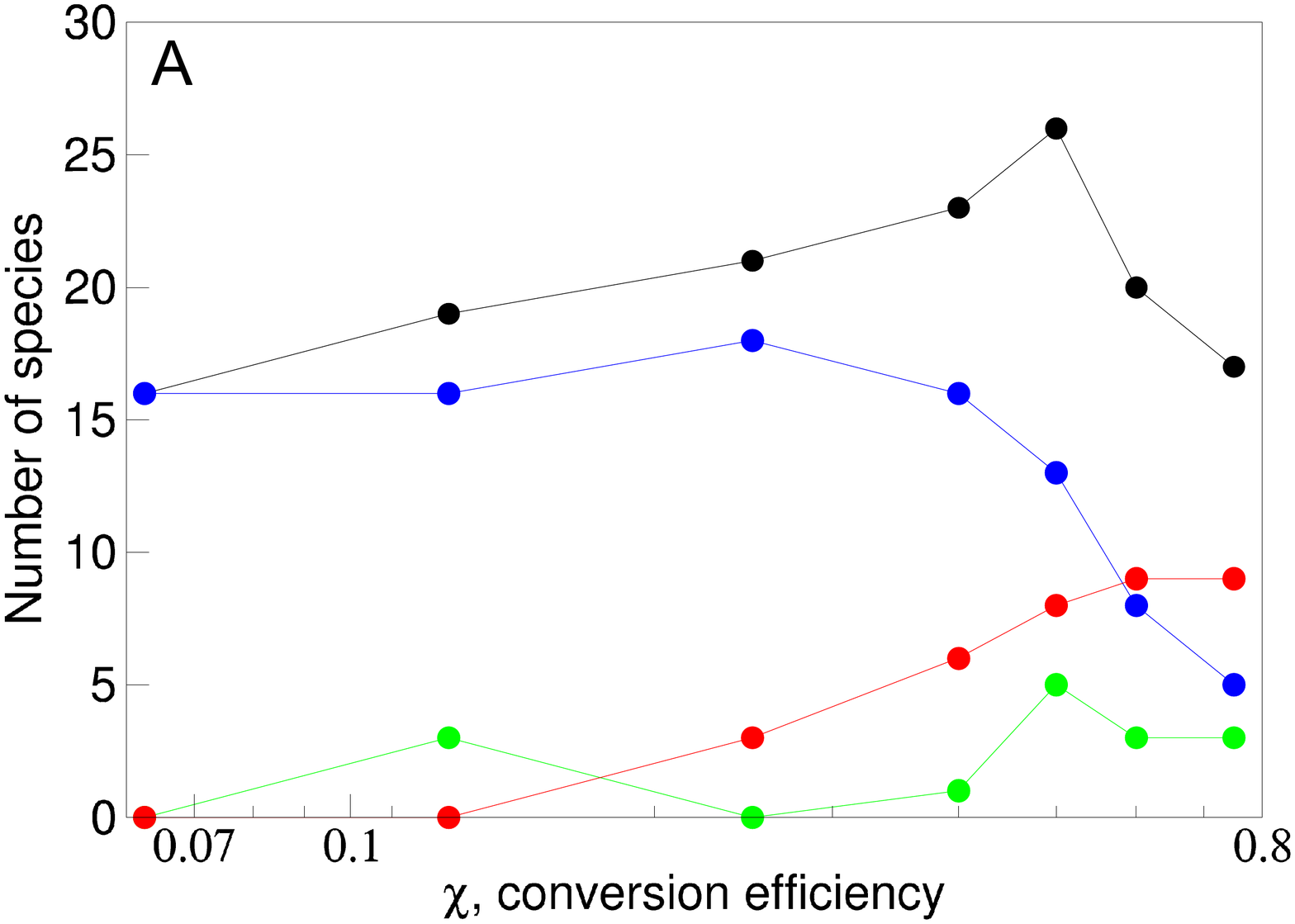}
\includegraphics[width=0.32\textwidth]{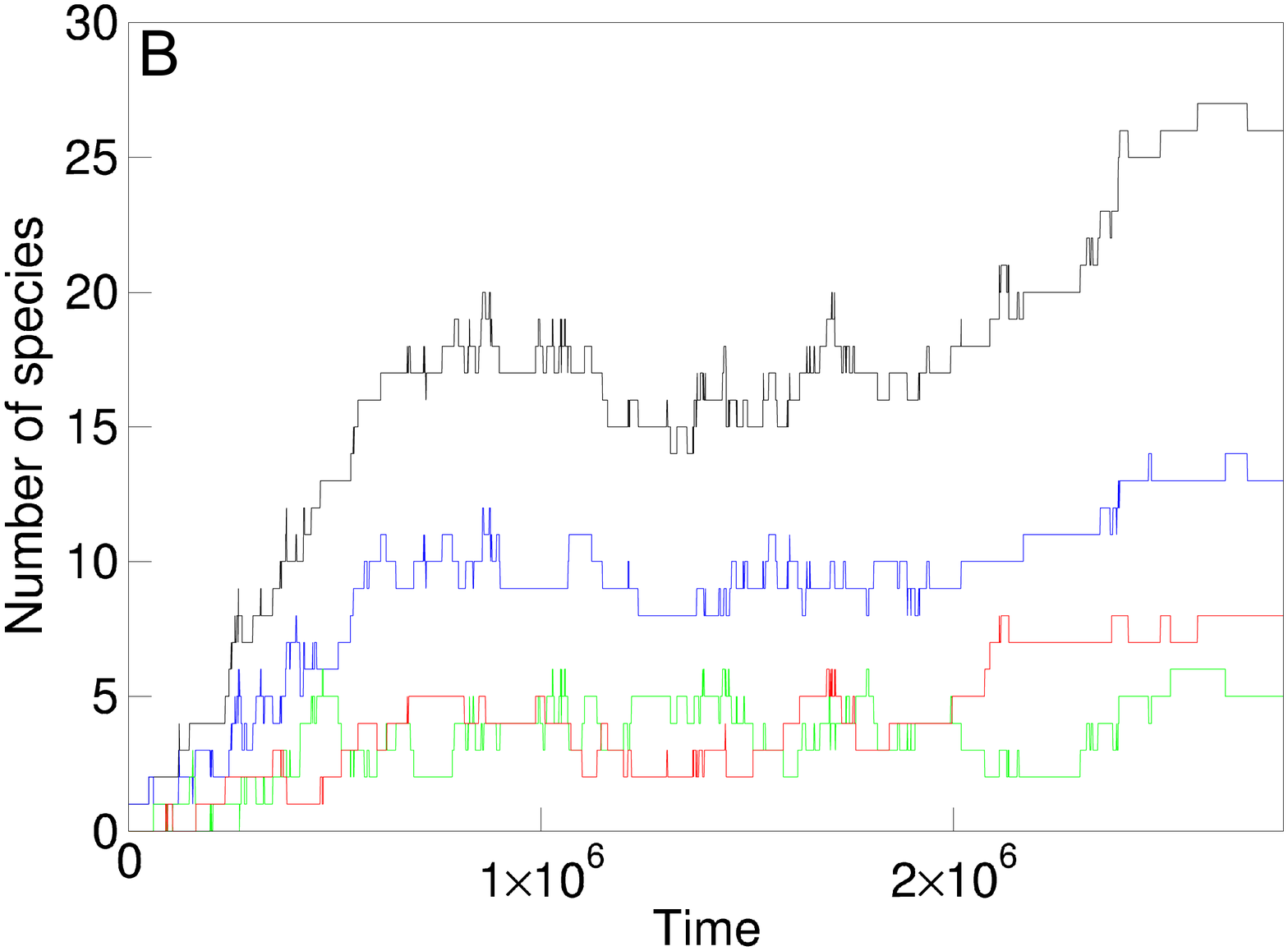}
\includegraphics[width=0.32\textwidth]{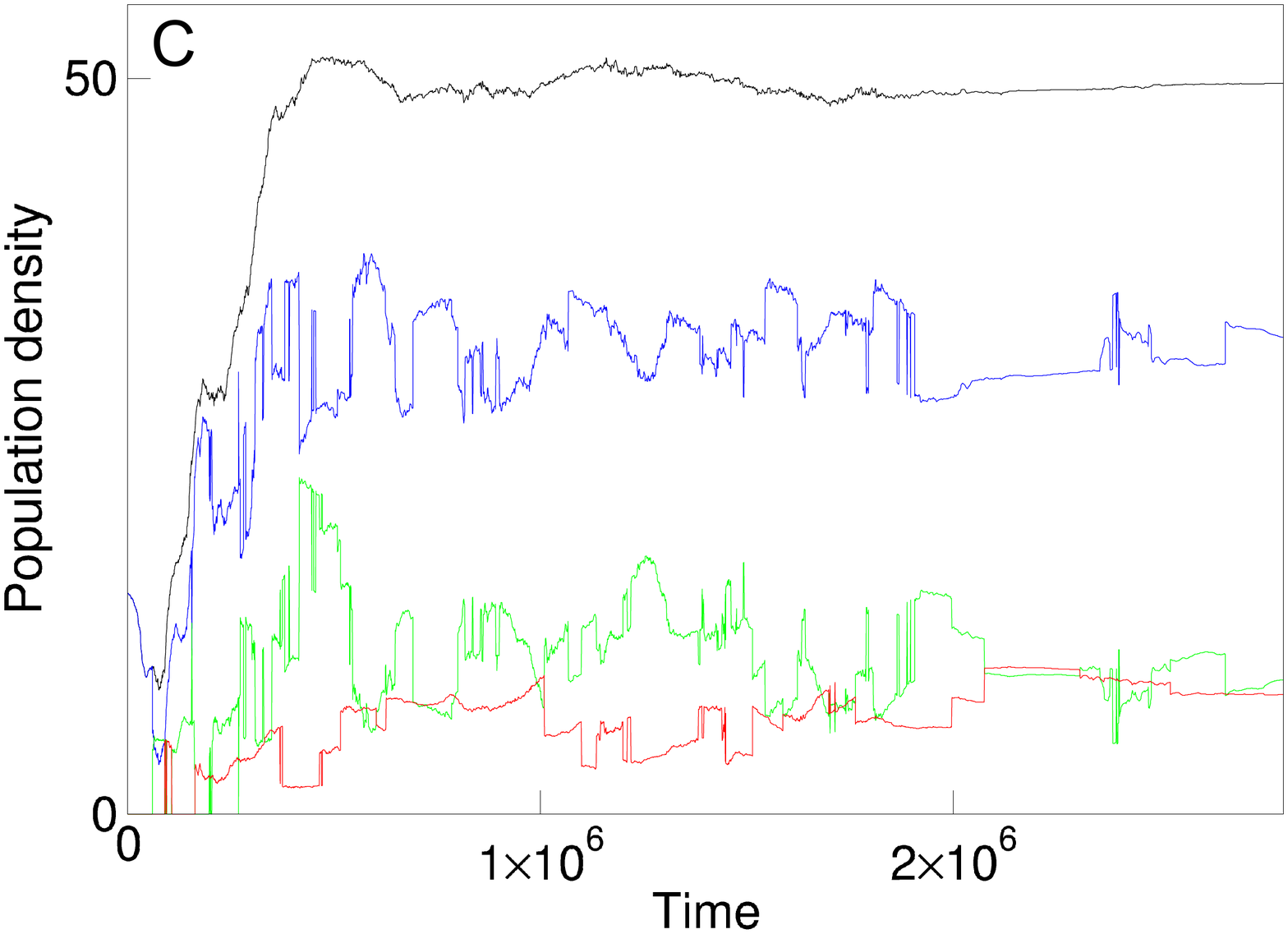}
		\caption{(a) 
                  The dependence of the steady-state levels of
                  diversity on the conversion efficiency $\chi$.
                   (b) The time dependence of the number of species
                   in a typical evolutionary scenario with $\chi=0.5$
                   (c) The time dependence of population densities in
                   the same scenario as in (b). The
                   total number of species in panels a) and b), and
                   the total population in panel c)
                   are shown by black lines, those quantities for
                   consumers are shown by
                   blue lines, for omnivores by green lines, and for predators
                   by red lines.} 
		  \label{f2}
\end{figure}

\subsection{Nonlinearity in tradeoff between resource consumption
  and predation affects diversity of omnivores}
In the majority of scenarios considered above, we observed that the
steady state community contained omnivore species, which both consume
primary resources and engage in predation. Intuitively, one would expect that the persistence of
such omnivore species may be conditional on the form of the tradeoff between the
two mechanisms of feeding: a more permissive tradeoff for combining both
mechanism ($\l>1$) may lead to more omnivores, while a more mutually
exclusive tradeoff ($\l<1$) may reduce or even completely rule out omnivores.
So far we considered a linear tradeoff, (\ref{tr}) with $\l=1$, and it is known that 
non-linear tradeoffs between consumption of various types of resources can in principle have dramatical effects on evolutionary outcomes
\cite{caetano2021evolution}.

\begin{figure}
             \centering
             \includegraphics[width=0.32\textwidth]{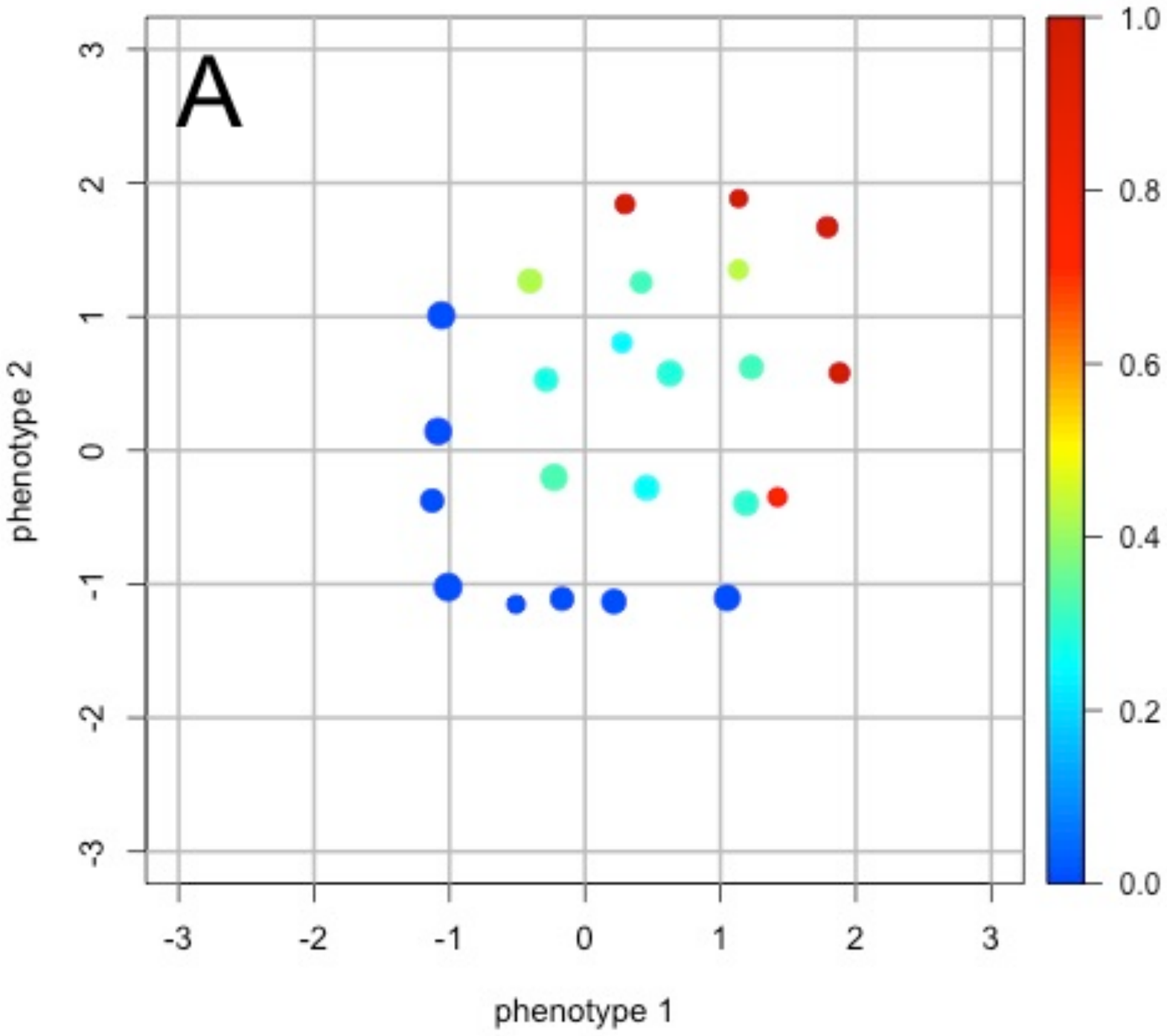}
             \includegraphics[width=0.32\textwidth]{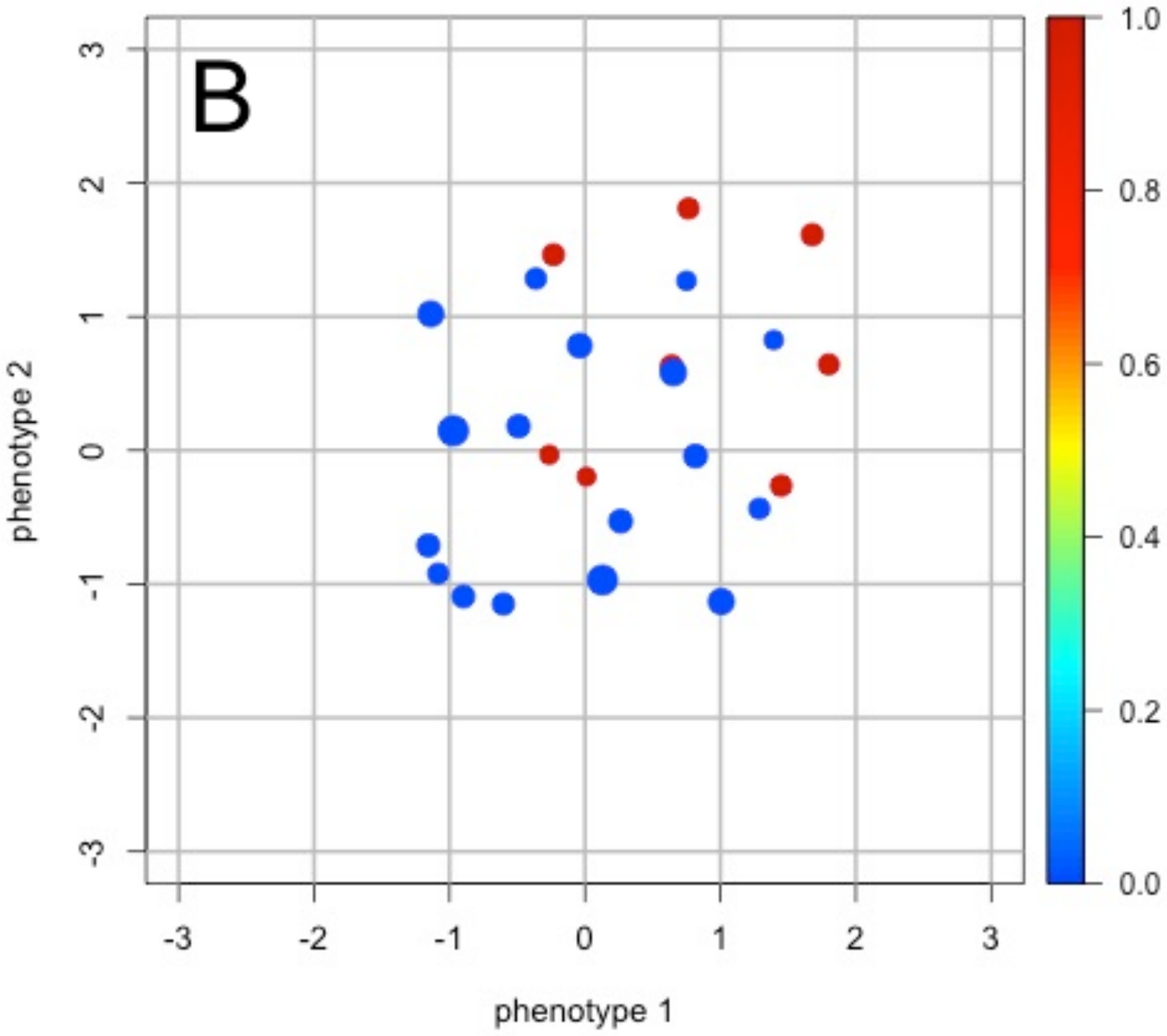}
             \includegraphics[width=0.32\textwidth]{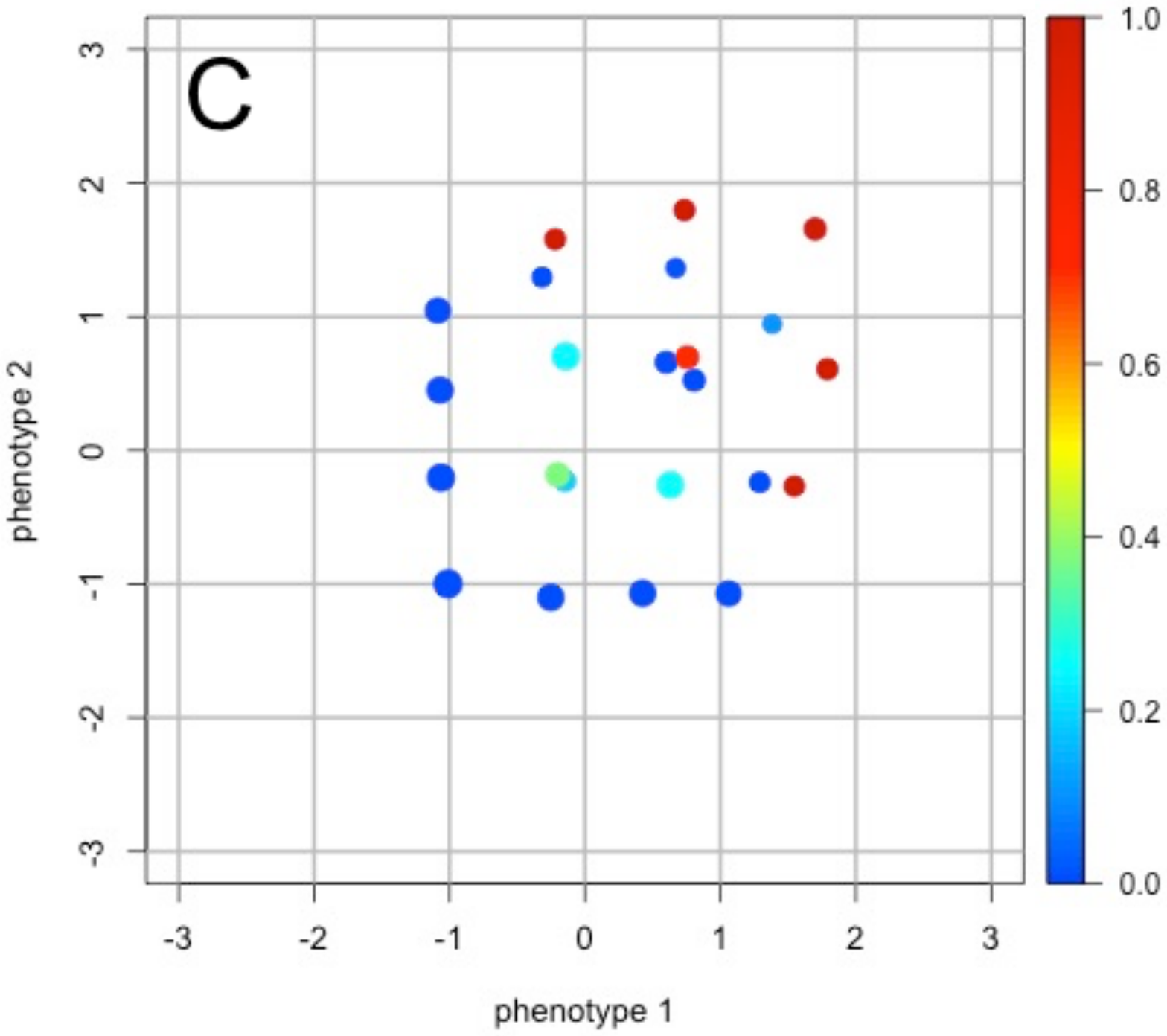}
 	      \caption{ 
                  Snapshots of steady state species distributions in
                   communities with   (a) convex tradeoff, $\l=1.1$,
                   (b) concave tradeoff, $\l=0.9$, and (c) linear
                   tradeoff, $\l=1$. In all three panels $\chi=0.4$
                   Videos of the evolutionary processes that led to these 
                  configurations can be found 
                  \href{https://figshare.com/s/1970d200680990011a96}{\underline{here}}. }
		\label{f3}
              \end{figure}

Therefore we explore the consequences of slight deviations of the tradeoff
exponent $\l$ defined in (\ref{tr}) from 1 in both the ``convex'' and
the "concave" direction.
As shown in Fig. \ref{f3}a, a convex or superlinear 
form of tradeoff with $\l=1.1$ preserves the overall distribution of species in the $x$-phenotype space, but results in several consumers and predators turning into
omnivores, i.e., in more intermediate species in the $p$-direction.
In contrast, a 
concave or sublinear tradeoff, $\l=0.9$  produces  a
more noticeable phenotypic rearrangement of species and 
completely eliminates omnivores, as shown in Fig. \ref{f3}b.

It follows that the very existence
of omnivores, as well as their fraction among all species in the ecosystem are sensitive to the type of trophic tradeoff assumed. In particular, robust presence of omnivory requires linear or superlinear tradeoffs $\l\geq1$. 

\subsection{Richer environments result in more predators and trophic levels}
We investigated how the richness of the environment, defined in
 our model by the coefficient $K_0$ in the carrying capacity, affects
 the evolving community. Here again, a common intuition (see
 e.g.\cite{takimoto2013environmental}) proves to be
 correct: a richer environment supports more species and longer
 food chain, while a scarcity of resources limits the
 diversity.

\begin{figure}
             \centering
             \includegraphics[width=0.32\textwidth]{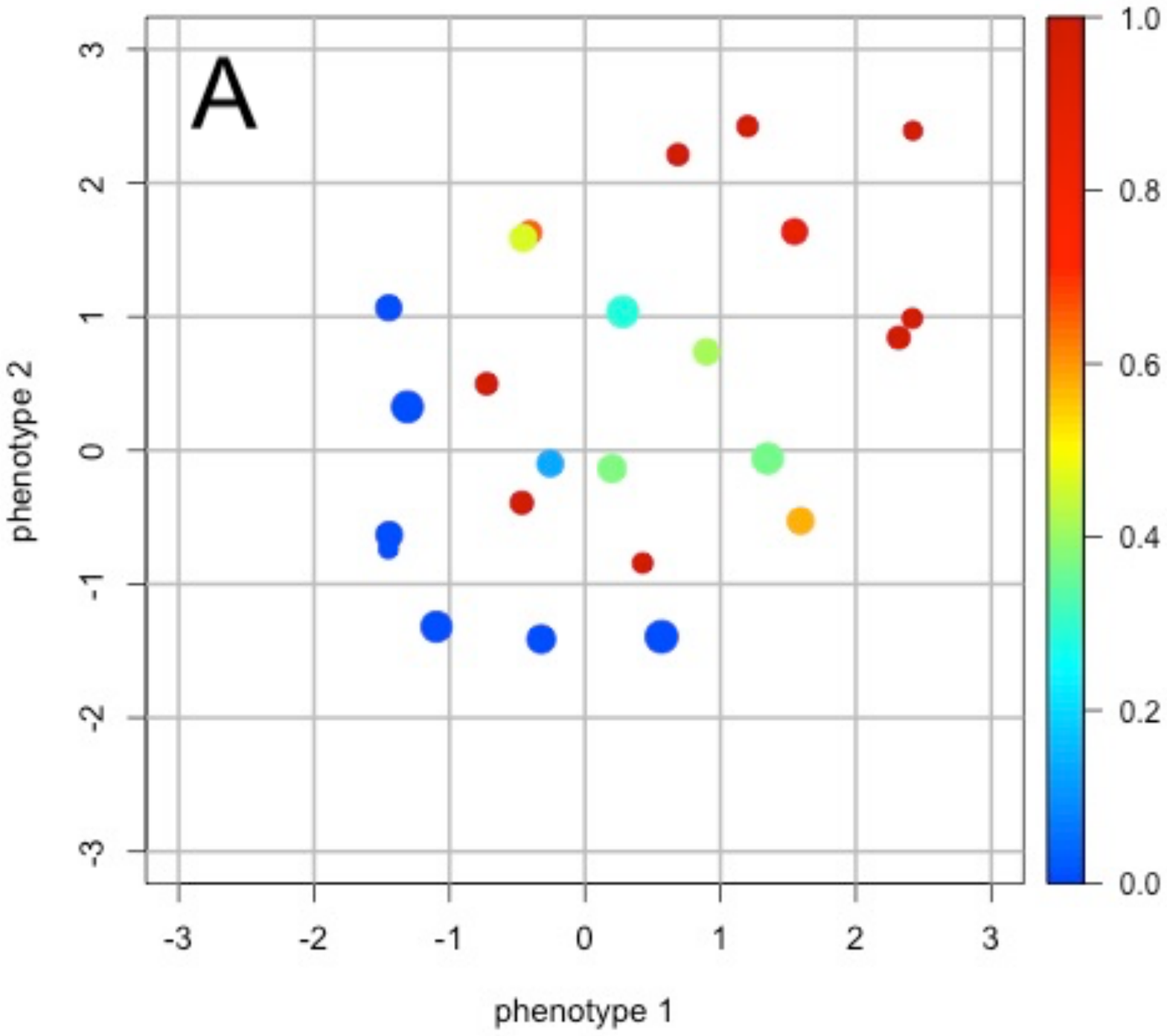}
             \includegraphics[width=0.32\textwidth]{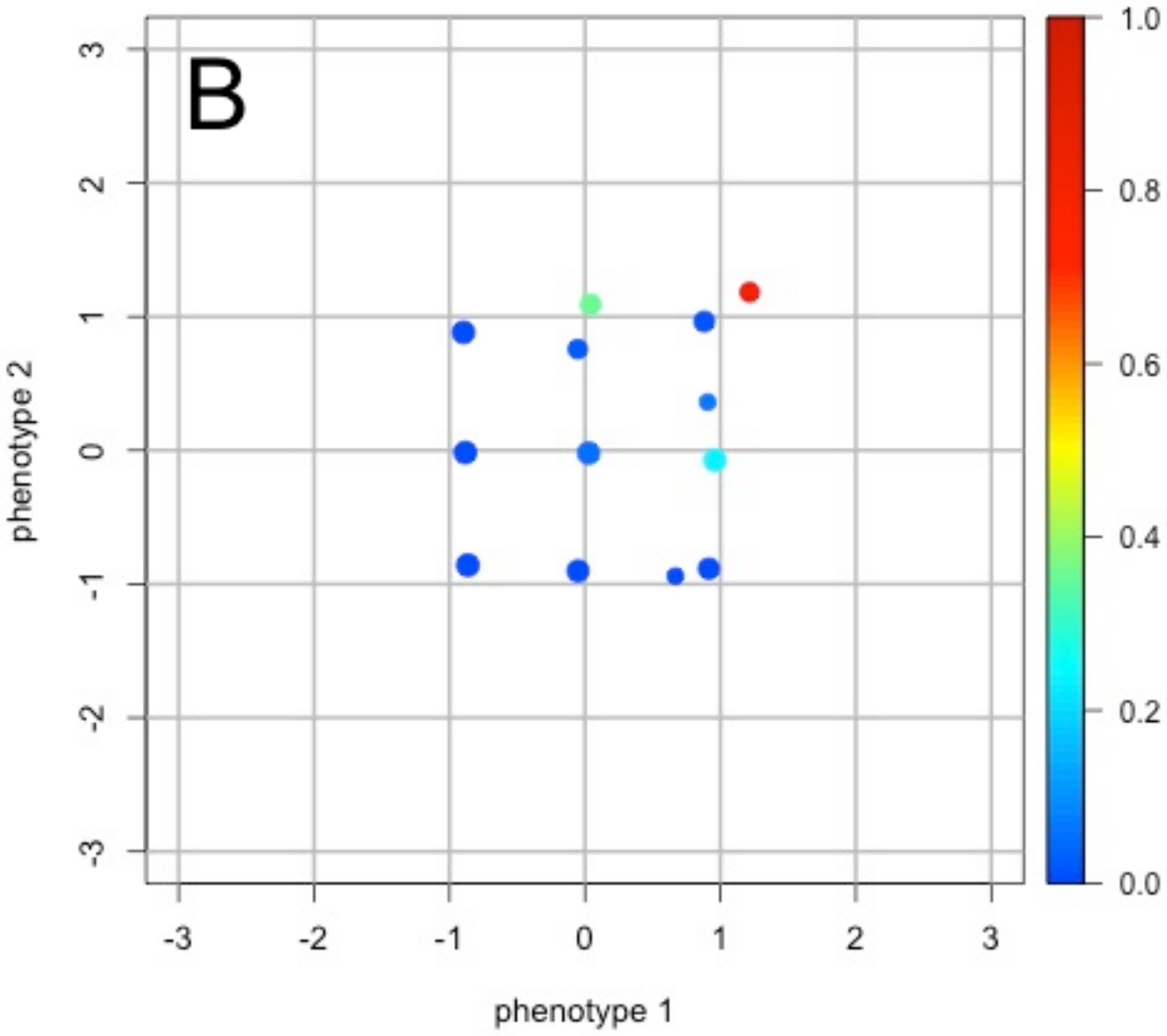}
             \includegraphics[width=0.32\textwidth]{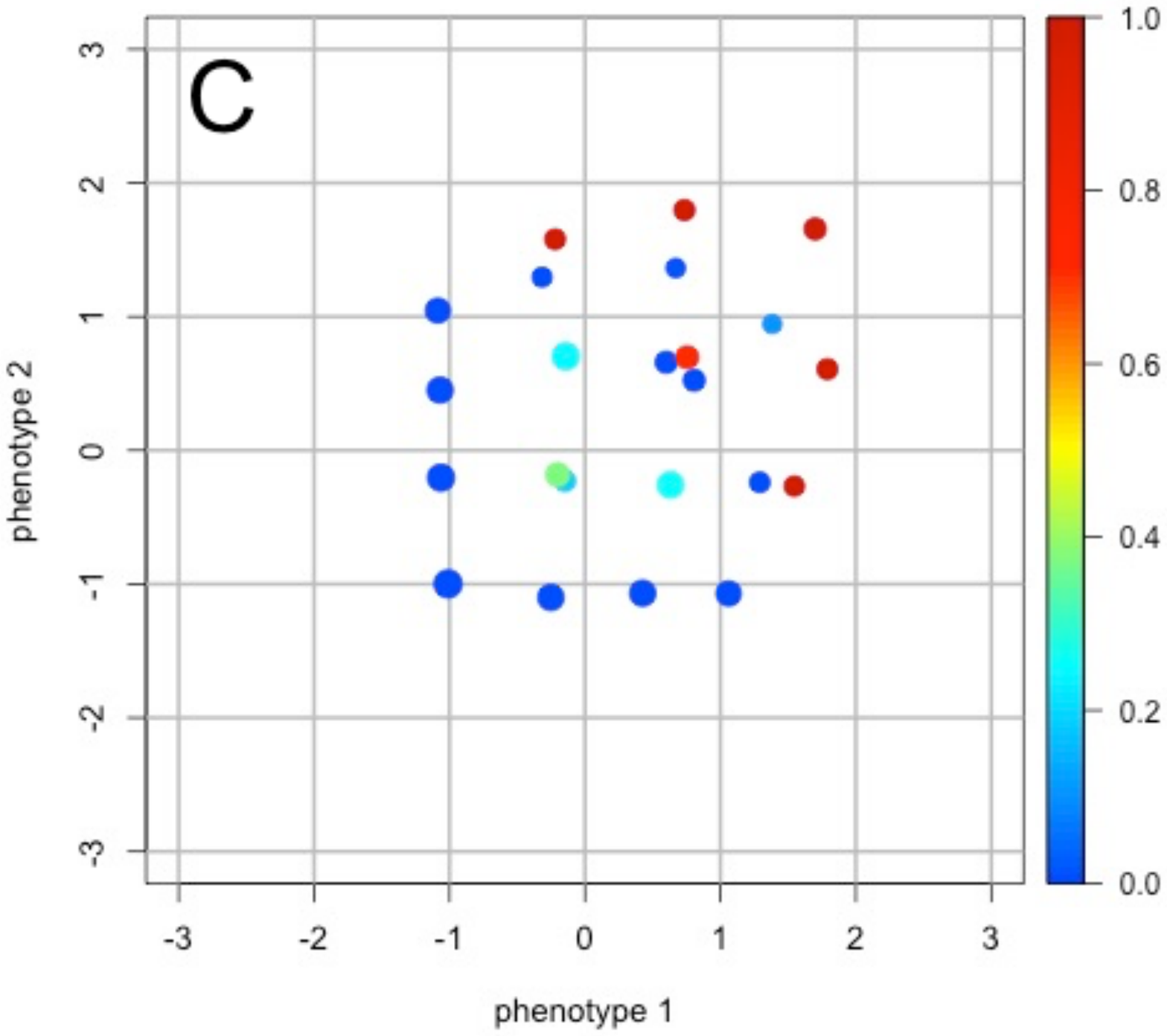}
 	      \caption{ 
                  Snapshots of steady state species distributions in
                   communities with  (a) rich environment, $K_0=16$,
                   (b) poor environment, $K_0=1$, and (c) reference
                   case $K_0=4$ as in Fig. 1d.
                   In all three panels $\chi=0.5$.
                   Videos of the evolutionary processes that led to these 
                  configurations can be found 
                  \href{https://figshare.com/s/1970d200680990011a96}{\underline{here}}. }
		\label{f4}
              \end{figure}

 Fig. \ref{f4} shows the steady state distribution of species 
 for the carrying capacity coefficient $K_0=16$, a 4-fold increase
 with respect to the reference level shown in Fig. 1d.
The larger carrying capacity allows more predator species
to evolve. Some of those additional species, visible near the center
of the phenotype space, feed on consumers that at lower carrying capacity were
not populous enough to support predation. Other additional predators
form higher trophic levels, preying on lower-level predators
rather than consumers.  Overall, an increase in the carrying capacity
tends to result in a reduction in the number of consumers and an increase in
the number of predators.

Fig. \ref{f4}b shows the distribution of species for a carrying
capacity  that is lower than in the reference case shown in Fig. 1d, $K_0=1$. Only a single predator, which feeds on
several consumers, evolves. It is interesting to compare this figure to the
scenario with the ``benchmark'' $K_0=4$ but a smaller $\chi$,
which also shows a single omnivore feeding on several
consumers Fig. \ref{f1}b. It appears that in a poor environment, the
single predator species depletes several consumers, so the consumer
diversity becomes noticeably smaller than in the case shown in
Fig. \ref{f1}b. 

Evidently, when the stochasticity of birth and death events is ignored
as it is done here, a reduction in the death rate produces the same effect as an increase in the birth rate. Thus we do not present results
for variation in the birth rate  $\b$, as it produces
changes similar to those caused by the variation of $K_0$ .

\subsection{The effect of the optimal offset between predator and prey
  phenotypes}
In all previous scenarios the offset $ \mathbf {m}$, defining the difference between predator and prey phenotypes in the $\mathbf{x}$-plane that maximizes the predation rate (see eq. (\ref{Gauss})) was set to $0.5$ in both the $x_1$
and the $x_2$ components. The very existence of such a (positive) offset is a
cornerstone of the models considered in \cite{loeuille2005evolutionary,allhoff2016biodiversity,brannstrom2011emergence,
  pillai2011metacommunity,bolchoun2017spatial,girardot2020does,
  fritsch2021identifying}, as it is a necessary
condition for distinguishing between prey and predator species and for the
generation of foodwebs in models that assume "automatic predation" when species have sufficiently different $\mathbf{x}$-phenotypes. However, as there are many examples of prey being subject to predation from
species with similar or smaller body sizes, and because in general, the phenotype $\mathbf{x}$ comprises not only body size, but potentially many other characteristics, 
 one should consider cases where $m$ can be either close to 0, or substantially negative as well as positive. 
Our model, in which the tendency to be a predator is a separate phenotype $p$, allows us
to meaningfully analyze such cases, including that when the offset is absent. We note that due to the symmetry of our models, cases with negative $m$ yield the same qualitative results as the corresponding cases with a positive $m$ of the same absolute value. In particular, examples in which predators evolve that are ``smaller'' than their prey can readily be produced. 
%i.e. $m=0$. shown in Fig. \ref{f5}.

\begin{figure}
             \centering
             \includegraphics[width=0.32\textwidth]{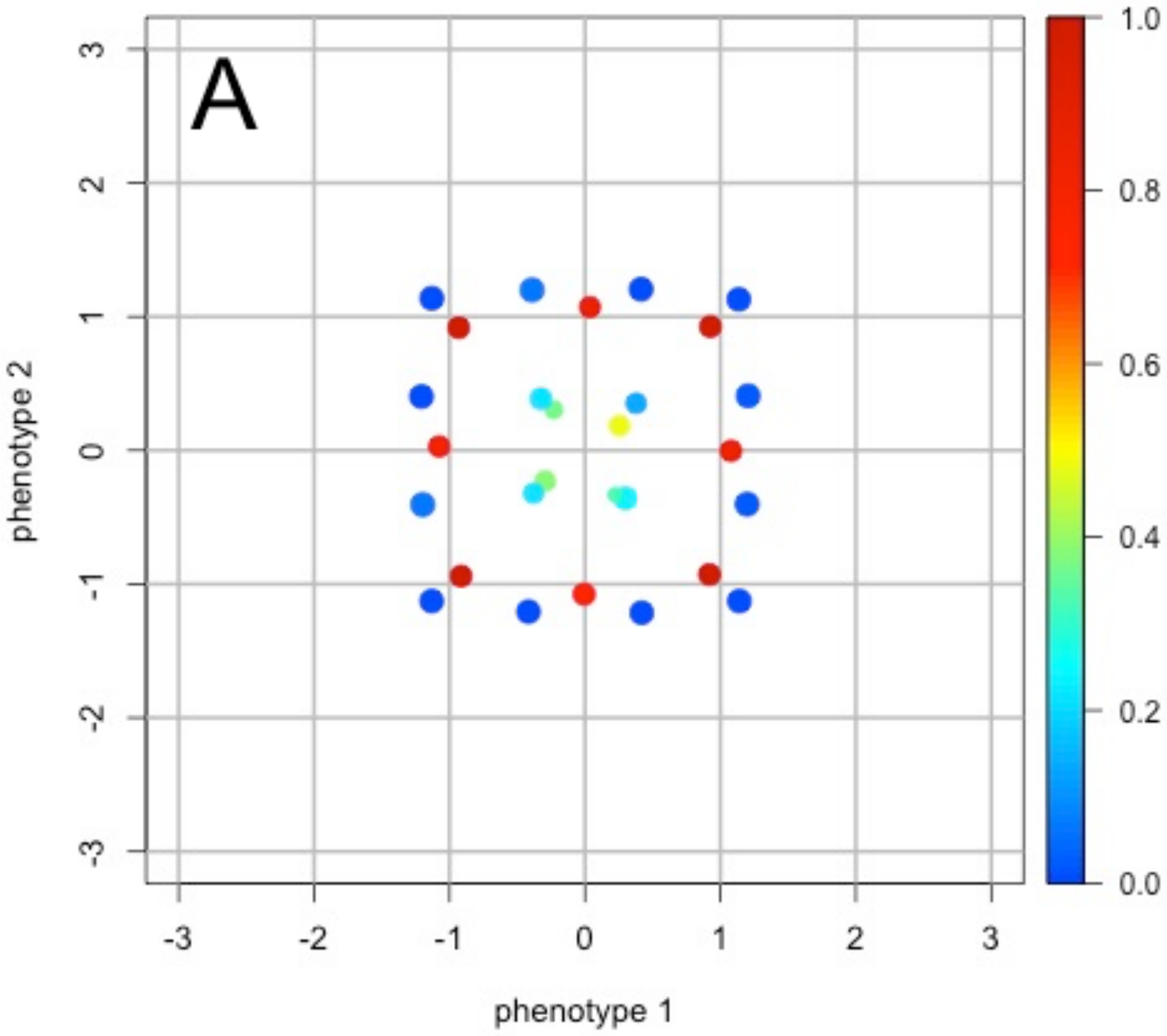}
             \includegraphics[width=0.32\textwidth]{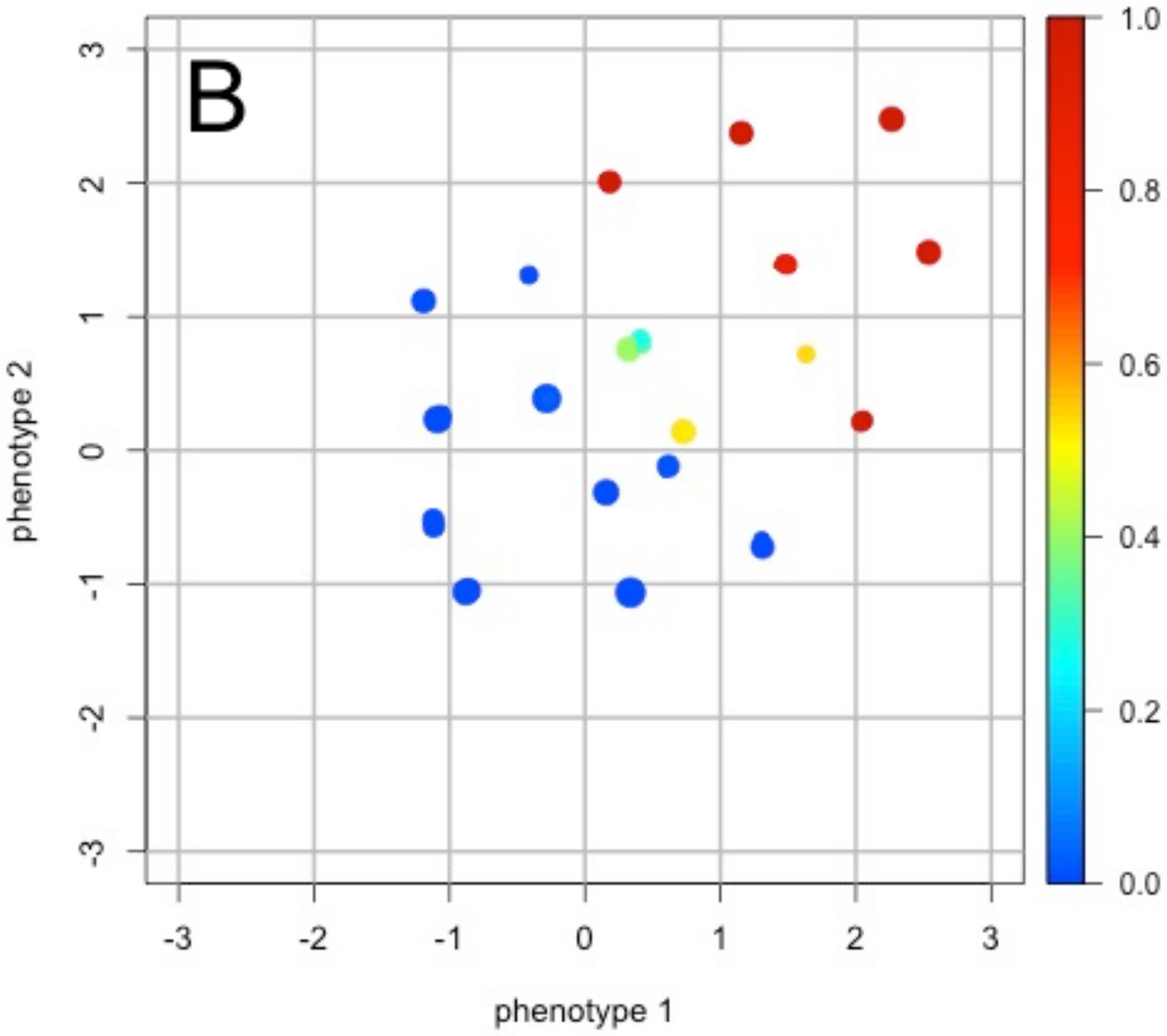}
             \includegraphics[width=0.32\textwidth]{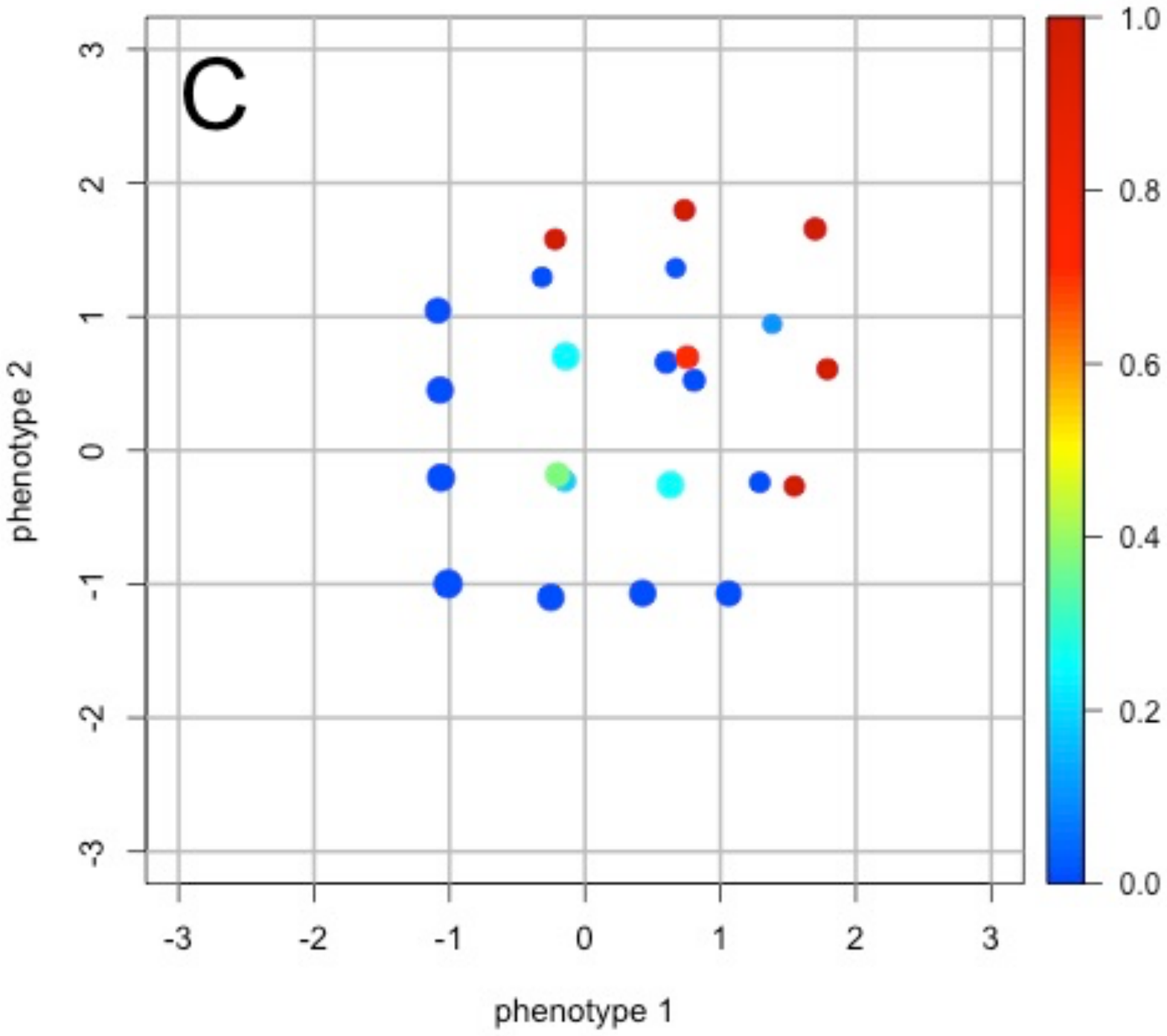}
 	      \caption{ 
                  Snapshots of steady state species distributions in 
                   communities with the optimal predation offset
                   (a) $m=0$, (b) $m=1$, and (c) $m=0.5$.
                   Videos of the evolutionary processes that led to these 
                  configurations can be found at
                  \href{https://figshare.com/s/1970d200680990011a96}{\underline{here}}. }
		\label{f5}
              \end{figure}

Fig. \ref{f5} shows that when $m=0$ (no offset), the consumer community evolves
to a phenotypic distribution that is very similar to the case with no
predation, Fig. \ref{f1}a. A similarly regular 
configuration of predators evolves phenotypically close to the
consumer to optimally prey on them.  Consumers that are phenotypically
more remote from predators (near the center of the figure) evolve a
certain degree of omnivory. Note that without the phenotypic offset,
higher-level predators do not evolve, which makes the resulting food
web rather flat. This is because once the first-level
predators appear, the optimal phenotypes of the next-level predators
would have been similar to those of the first-level ones, making them
indistinguishable. 

An offset of $m=1$ (larger than the reference offset) results in food chains with
larger phenotypic separation between levels, Fig. \ref{f1}b.
The phenotypes of all predators become noticeably larger than those of any
prey, and the maximum prey phenotype exceed those in the reference
system, Fig. 5c.

\section{Discussion}
We presented a model for evolving ecosystems  based on competition for
primary resources and allowing for the evolution of predation on other species in the
community. The ability to be a predator is  evolving independently of the phenotypic features that determine the strength of ecological interactions. Both competition for primary resources and predation attack rates are determined by the same set of relevant morphological, physiological and behavioural traits, which is described by a generally multi-dimensional, quantitative phenotype vector. Competition is defined by the competition kernel, for which the simplest (and classical) assumption is that it has a maximum when the phenotypes of competing individuals are the same. For the attack kernel, the simplest assumption is that it has a maximum at a given difference between the phenotypes of the interacting predator and prey individuals. In addition, we assumed that the predation and
resource consumption abilities are mutually restrictive
through a tradeoff. The model reproduces many 
known relevant features of emergent food webs and makes some
predictions that were not emphasized earlier.

In particular, we confirm the observations
\cite{fritsch2021identifying} that the conversion efficiency plays a
key role for the evolution of predation and the structure of the resulting
food chains. Low conversion efficiencies tend to exclude the emergence
of predation, and in that case, evolution results in a diverse
community of 
primary resource consumers. A more efficient conversion allows one or a few
predators to evolve, which appear only after the consumers 
have sufficiently diversified to provide enough prey. An even higher conversion
efficiency produces an evolutionary increase of predatory
abilities even when the consumer community consists of a single species.

The overall diversity of consumers and predators in the evolving
ecosystem reaches a maximum at intermediate conversion 
efficiencies, while systems with lower or higher conversion efficiency evolve
either fewer predators or fewer consumers. In many cases,
multi-trophic food webs emerge, containing several omnivore species. The richness of the environment,
expressed as the scale of the carrying capacity, plays a role that is similar 
to that of  the conversion efficiency, albeit simultaneously affecting both the
consumers and predators.

A  previously unexplored scenario occurs when the attack rate is maximal for very similar predator and prey phenotypes. 
In that case, the resulting food web consists of only two
trophic levels, 
with the first-level predators preying on consumer but lacking any
higher-level predation. The model also shows many ecological scenarios
where predators feed on prey that are larger, which in nature occurs at many size size scales, from bacterial predators attacking other bacteria
\cite{hungate2021functional} to
large terrestrial
predators preying on equal-sized or even larger herbivores
\cite{john2009lion}. The latter 
systems tend to exhibit short food chains, which is  similar to
what we observe.  
In the complementary scenario, when predation on
prey with much smaller phenotypes is optimal and the conversion efficiency is high,  the
model predicts development of multilevel food chains similar to
those observed in aquatic environment. 

Under a broad set of conditions the system also evolves omnivory,
defined as intermediately-developed abilities to consume resources and
prey. However, the evolution of omnivory strongly depends on the shape
of the tradeoff between competitive and predatory
capabilities. Specifically, a concave (sublinear) relationship,
favouring specialist consumers or predators  over omnivores,  does prevent
the evolution of omnivory, which requires a linear or convex
(superlinear) tradeoff. This is in a complete accordance with the
very broadly applicable
conclusion that a convex tradeoff favours generalists and concave 
tradeoff results in more specialists \cite{gonzalez2022modeling, caetano2021evolution}. 
Given that the biochemical, 
morphological and behavioural contributions to this
tradeoff are complicated in general, it is difficult to conjecture
what the shape of the tradeoff would be in real systems (except to say
that exact linearity may be very rare).  

We hope that our work sets up a more comprehensive framework for analyzing
evolution of predation. It effectively opens up a new and independent
dimension to this problem by uncoupling the evolution of predatory
abilities from  evolution of other  phenotypes that determine the
strength of ecological interactions, such as body size, but also many
other physiological and morphological traits. 
Here we report an initial schematic implementation of this
framework that nevertheless makes several realistic predictions and
confirms existing observations. Furthermore, the suggested framework seems easily
adaptable to more complex evolutionary scenarios:  For example, for
certain classes of phenotypic coordinates it
would be realistic to incorporate the allometric scaling of rate
constants \cite{yodzis1992body} as it is done, for example, in
\cite{loeuille2005evolutionary,  fritsch2021identifying}.  Making the
consumption of resources explicit rather than implicitly describing it using
logistic equations is another possible extension. The role of the
dimensionality of phenotype space, grossly understudied in all
existing models due to computational complexity, also remains to be elucidated
in the context of our model. Finally, since conversion efficiency
plays a crucial role for model dynamics, it would be interesting to
investigate the evolution of the conversion efficiency and its
dependence on other traits.

Overall, we think that considering predation ability as an independently
evolving property is realistic, and therefore a useful extension of
existing work. Our models present a first step in that direction.

%\section*{Conclusion}

\section*{Acknowledgments}
MD was supported by NSERC Discovery Grant 219930. YI acknowledges support from FONDECYT project 1200708.

\appendix
%\paragraph*{S1 Appendix.}
\label{S1_Appendix}
\section{An Adaptive dynamics estimate for the
  minimum conversion efficiency to evolve predation}
Here we evaluate the conditions
under which a single species starts to
 develop predation abilities, that is, to evolve towards non-zero
 $p$. A  description of the adaptive dynamics approximation, that
 is used in the following, can be found, for example, in
 \cite{dieckmann1996dynamical, doebeli2011adaptive}).

 Consider the invasion fitness, i.e., the per capita growth rate, of a rare
mutant with coordinates $(\mathbf{y},q)$ in a system populated by a single
species that is monomorphic for phenotype  $(\mathbf{x},p)$ and has population size $N$,
\begin{align}
\label{a1}
  f(\mathbf{y},q;\mathbf{x},p)=(1-q)\beta -
  \delta-(1-q)(1-p)N\frac{\a(\mathbf{y,x})}{K(\mathbf{y})}+\\
  \nonumber
  +N\left[ q \chi \g(\mathbf{y,x}) - p \g(\mathbf{x,y}) \right].
\end{align}
Here we assumed a linear tradeoff between competitive and predatory ability, that is, $\l=1$ in (\ref{tr}).
The evolution of $p$ is proportional to the corresponding component of the
selection gradient,
\begin{align}
\label{a2}
\frac{dp}{dt} \propto s_p\equiv \left.\frac{\partial f}{\partial q}\right\vert_{\mathbf{y}=\mathbf{x},
  q=p}
  = - \beta + N \left[ (1-p) \frac{\a(\mathbf{x,x})}{K(\mathbf{x})} + \chi \g(\mathbf{x,x}) \right]
\end{align}
The factor $1-p$ in the second term ensures that if the selection
gradient in $p$ is negative for $p=0$, it will remain negative for
larger $p$. Hence, we evaluate the threshold values for the parameters
when the right hand side
of Eq.(\ref{a2}) is negative. 

For the steady state population of a single
species with $p=0$, Eq. (\ref{e1}) yields
\begin{align}
\label{a3}
N=K(\mathbf{x})\frac{\b-\d}{\a(\mathbf{x,x})},
\end{align}
so that the selection gradient becomes
\begin{align}
\label{a4}
s_p=-\d + N\chi\g(\mathbf{x,x})=-\d+\chi (\b-\d)\frac{\g(\mathbf{x,x})}{\a(\mathbf{x,x})}K(\mathbf{x}).
\end{align}
This means that for a single species at $x=0$ and with the functional forms of 
$K(\mathbf{x})$, $\a$ and $\g$ chosen as in the main text, evolution of $p$ away from 0 is possible when
\begin{align}
\label{a5}
\chi (\b-\d) \s_{\a} ^2\exp\left(-\frac{m^2}{\sigma_{\g}^2}  \right)K_0>
  \d\s_{\g}^2.
\end{align}
For the ``reference'' case in the main text, this estimate gives
$\chi>0.23$. This qualitatively agrees  with 
 the numerical observations summarized in Fig. 1, which shows that the
 transition between initial diversification of consumers to the
 immediate increase in $p$ of the single initially introduced species
 happens for $0.25<\chi<0.4$.  


 \bibliographystyle{elsarticle-num} 

%\bibliography{predation}

\end{document}